\documentstyle[12pt,bezier]{article}

\let\a=\alpha
\let\b=\beta
\let\g=\gamma
\let\d=\delta
\let\e=\epsilon

\let\h=\eta

\let\l=\lambda
\let\m=\mu
\let\n=\nu

\def\p{\vec{p}}
\def\q{\vec{q}}
\let\r=\rho
\let\s=\sigma

\let\o=\omega

\let\S=\Sigma

\let\L=\Lambda
\let\G=\Gamma
\let\D=\Delta

\let\u=\underline

\def\2{{1\over2}} \def\4{{1\over4}} \def\52{{5\over2}}
\def\6{\partial }

\def\({\left(} \def\){\right)} \def\<{\langle } \def\>{\rangle }
  \def\lb{\left\{} \def\rb{\right\}}

\def\CD{{\cal D}}

\def\CV{{\cal V}}

\def\beg{\begin{equation}}
\def\begar{\begin{eqnarray}}
\def\ee{\end{equation}}
\def\ea{\end{eqnarray}}

\newcommand{\pref}[1]{(\ref{#1})}                
\newcommand{\plabel}[1]{\label{#1}}              
\newcommand{\prefeq}[1]{Eq.~(\ref{#1} )}         
\newcommand{\pcite}[1]{\cite{#1}}                
\newcommand{\pbib}[1]{\bibitem{#1}}              

\begin{document}


\begin{titlepage}
\hfill TUW-94-14
 \begin{center}
\Large
 {\bf Rigorous QCD-Potential for the $t\bar{t}$-System at
Threshold}\\[1cm]
\normalsize
 W. Kummer and W. M\"odritsch$^*$\\[1.2cm]
 Institut f\"ur Theoretische Physik, Techn. Univ. Wien\\
 A-1040 Wien, Wiedner Hauptstra\ss e 8-10
\end{center}
\vspace{1.0cm}
\centerline{Abstract}

\begin{sloppypar}
Recent evidence for the top mass in the region of 160 $GeV$
for the first time provides an opportunity to use the full power
of relativistic quantum field theoretical methods, available also
for weakly bound systems. Because of the large decay width $\G$
of the top quark individual energy-levels in "toponium" will be
unobservable. However, the potential for the $t\bar{t}$ system,
based on a systematic expansion in powers of the strong coupling
constant $\a_s$ can be rigorously derived from QCD and plays a
central role in the threshold region. It is essential that the
neglect of nonperturbative (confining) effects is fully justified
here for the first time to a large accuracy, also just {\it because}
of the large $\G$. The different contributions
to that potential are computed from real level corrections near the
bound state poles of the $t\bar{t}$-system which for $\G \ne 0$
move into the unphysical sheet of the complex energy plane.
Thus, in order to obtain the different contributions to
that potential we may use the level corrections at that (complex) pole.
Within the relevant level shifts we especially emphasize the corrections of
order $O(\a_s^4 m_t)$ and numerically comparable ones to that order
also from electroweak interactions which may become important as
well.
\end{sloppypar}

\vspace{\fill}

\_\hrulefill \hspace*{8cm} \\
Vienna, July 1994
$^*$ e-mail: wmoedrit@ecxph.tuwien.ac.at
\end{titlepage}

\section{Introduction}

Perturbative expansions in the coupling constant in quantum field theory
possess two types of applications, the calculation of scattering processes and
the computation of processes involving weakly bound systems. Many of the
successes of quantum electrodynamics (QED) are, in fact, related to
positronium, i.e. to the second one of the aforementioned applications. The
proper starting point for any bound-state calculation in quantum field theory
is an integral equation, comprising an infinite sum of Feynman graphs. The
Bethe-Salpeter (BS) equation
\pcite{Bethe} fulfills this task, and it is well known that in the limit of
binding energies of $O(\a^2m)$ the Schr\"odinger equation with static Coulomb
attraction is obtained. The computation of higher order corrections to the
Bohr levels - beyond $O(\a^5 m)-$, however, turned out to be far from trivial.
It was recognized, though, relatively late that, at least conceptually,
substantial progress with respect to a systematic treatment results from a
consistent use of a perturbation theory geared to the original BS equation
\pcite{Lep77}. In that manner, at the same time, nonrelativistic expansions as
implied by Hamiltonian approaches with successive Fouldy-Wouthuysen
transformations \pcite{Fein} are avoided.

Within the BS-technique it is desirable to have an exactly solvable zero
order equation different from the Schr\"odinger equation, otherwise e.g. the
approximation procedure for the wave function lacks sufficient transparency,
especially in higher orders.  Considerable freedom exists for selecting such
an equation. Of course, certain corrections included already at the zero level
are to be properly subtracted out in higher orders. An especially useful zero
order equation has been discovered some time ago by Barbieri and Remiddi (BR
equation \pcite{BR}), but also other equations have been proposed
\pcite{equs}. Still, one of the most annoying features of all bound-state
calculations remains the pivotal rule played by the Coulomb gauge. In other
gauges, e.g. already the (in QED vanishing) corrections $O(\a^3m)$ of the Bohr
levels imply the inclusion of an infinite set of Feynman graphs \pcite{Love},
already at that order. Only in very special cases, when certain subsets of
graphs can be shown to represent together a gauge-independent correction,
another more suitable (covariant) gauge may be chosen.

In contrast to positronium in QED, the vast literature on bound state problems
in quantum chromodynamics (QCD) adheres to a description of the
quark-antiquark system by the Schr\"odinger equation with corrections
'motivated' by QCD \pcite{ph}. As long as a relatively small number of
parameters suffices for an adequate phenomenological description of observed
quantum levels, this approach doubtless has an ample practical justification.
However, again and again certain deviations from such phenomenological
descriptions are reported
\pcite{Franz}.  Thus also for this reason a return to more rigorous QCD
arguments remains as desirable as ever - provided Nature offers a "window"
where such methods are applicable. The standard literature on this subject
almost exclusively is based on nonrelativistic expansions \pcite{Fein} or on
the calculation of purely static Coulomb forces \pcite{Fisch}. Also very often
potentials with higher order corrections as determined from on-shell quarkonia
scattering are used \pcite{scatt}.  A comprehensive account of this usual
approach is given in \pcite{Buchm}.  However, one cannot hope to push some of
the calculations further without a solid theoretical basis {\it which is
provided by the Bethe-Salpeter formalism alone}. Terms not expressible from
the outset as "potentials", or relevant off-shell effects, which are typical
for higher order corrections, may be even lost  \pcite{Bkomm}.  The importance
of such a perturbatively consistent approach is supported by the empirical
fact that at least in one case, namely the decay of S-wave quarkonium, the
result of a full BS-perturbation calculation \pcite{KumW}, including the QCD
corrections to the bound state wave function, yields a result very different
from the one which considered only the corrections to the quark antiquark
annihilation alone
\pcite{BC}.

In this connection the relatively large running coupling constant even at high
energies represents a well known problem, together with the frequent encounter
of large coefficients in a perturbative expansion. For this reason e.g.
problems arise in the comparison of the coupling constant as determined from
scattering experiments within the minimal subtraction scheme
($\overline{MS}$), with the coupling constant to be used in a consistent weak
bound-state approach. The philosophy within our present work will be that the
orders of magnitude, as determined from $\a_{\overline{MS}}$, will be used for
estimates but that we shall imply a determination of $\a_s$ by some physical
observable of the quarkonium system itself (e.g. energy levels, cf. the
remarks after eq. \pref{pi1er} ).  In that way delicate correlations of
'genuine' orders of $\a_s$ from {\it basically} different types of experiments
are avoided.

Of course, the quarkonium system also differs profoundly from positronium
because of the confinement of quarks and gluons. However, the phenomenological
success of the nonrelativistic quarkonium model can be explained by the fact
that the bound states of heavy quarkonia are sufficiently deep in the Coulomb
funnel and thus sufficiently far away from the confinement part of the
potential.  Estimates in the early 80-s \pcite{Leut} of nonperturbative
effects, describing the 'tail' of confinement by a gluon condensate
\pcite{Shif} suggested that only with quark masses well above about 50$GeV$
the importance of such nonperturbative effects for low Bohr quantum numbers
may decrease sufficiently to make perturbative 'field theoretical' level
corrections competitive and observable.  An improved calculation of that
correction  based already on the BS approach was done some time ago
\pcite{KumWL} and shows a result which is smaller by about 20\% as compared to
the estimate based upon the Schr\"odinger equation.  Now from CDF \pcite{CDF}
and from high precision electro-weak experiments of the LEP collaborations
\pcite{ALEPH}, the mass range of the top quark seems to be established to lie
in the range 150-180$GeV$.  Thus for the first time a nonabelian bound state
quarkonium system seems to fulfill the high mass requirement for a genuine
field theoretical approach.  On the other hand, a special feature of a
superheavy top is that the weak decay $t \to b+W$ broadens the energy levels
\pcite{Kuehn} for increasing mass $m_t$ so that above $m_t \approx 140 GeV$
individual levels effectively disappear.  This, however, by no means
invalidates the use of perturbation theory for weakly bound systems. E.g in a
future $e^+ e^-$ collider experiment at the $t\bar{t}$ threshold a smooth
curve - instead of sharp energy levels immediately below the threshold - will
the object of experimental studies. Nevertheless also  this {\it curve} can be
predicted theoretically.  Moreover, the large width $\G_{t \to b + W} >>
\L_{QCD}$ even turns into a virtue since it makes confining effects
unimportant: the top has no time to hadronize and thus becomes to a rather
high accuracy a fully perturbative object, even for QCD \pcite{Khoze,Peskin}.

In contrast to low mass quarkonia this also offers the possibility to extend
bound-state perturbation arguments to (nonrelativistic) energies $E > 0$ above
the $t\bar{t}$ threshold $E=0$.  As shown first in the seminal work of Fadin
and Khoze \pcite{Khoze} it is possible to include the width in the zero order
nonrelativistic Greenfunction by the simple replacement $E \to E + i \G$. The
Greenfunction $G$, whose absorptive part expresses the cross section for
$t\bar{t}$ production, beside $E + i \G$ only depends on the potential.
Previous work along this line \pcite{Khoze,Peskin,Sumino} was based on
QCD-suggested and/or phenomenological potentials. The aim of our present study
is to find a way to replace those potentials by one, derived from systematic
QCD perturbation theory only.  Any ambiguity in our approach is eliminated by
the requirement that this potential should be identified order by order with
the corresponding one for "nonabelian positronium", evaluated from (real)
energy shifts at the position of the bound-state singulatity which has moved
into the unphysical sheet of the complex energy plane. It should be noted that
well known arguments \pcite{Velt} for such an aproach even guarantee the
gauge-independence of our results, despite the fact that, for practical
reasons, the calculations are performed in the Coulomb gauge. Proofs of
gauge-independence are valid order by order in perturbation theory. Also for
that reason we are careful to keep track of these order, meticulously avoiding
e.g. the mixing of orders in a running coupling constant.

To $O(m\a_s^3)$ for the energy shifts one encounters the well-known
vacuum polarization effects, present in a system of two fermions with
equal masses $m$ amd gauge coupling $g=\sqrt{4 \pi \a_s}$, if other
fields with lower mass ( here the lighter flavours and the gluons )
are present.

At $O(m\a_s^4)$ the 'relativistic' corrections, corresponding to the same type
of graphs as in the abelian case are present here as well. However, even
considering only vertex corrections alone, a difference to the abelian case
was discovered a long time ago
\pcite{Dun}. In addition typical other nonabelian contributions may appear at
this level as well. In contrast to scattering processes, the determination of
the order to which a certain graph contributes in relativistic bound state
perturbation theory in each case requires a special analysis. Other terms
(from QED and weak interaction), incidentally, may be of the same {\it
numerical} order $O(m
\a_s^4)$.  The Coulomb gauge also entails peculiar additional nonlocal
interaction terms in the effective action, appearing e.g. in the path integral
formulation
\pcite{Schw}. We also investigate the effect of those terms here.

In section 2 we recall some basic facts about BS-perturbation theory.  The BR
equation for nonabelian weakly bound onium-systems is found to allow a simple
generalization for wave functions at complex poles. In fact the advantages of
the BR equation really become more evident above $O(\a_s^4)$.  However we want
to introduce this technique here, because it has shown to yield already at
least in one case also a very convenient regrouping of corrections: In a
computation of the bound state effects for the decay width of a "toponium"
state a Ward -like cancellation of graphs became evident most clearly just in
this approach \pcite{Moed}.

Tree graphs leading among others to the well known $O(m \a^4_s)$ corrections
are discussed in sect. 3. As indicated already above, the one loop vacuum
polarization (sect. 4.1) provides a correction term $O(m \a_s^3)$ in the
nonabelian case. Here we take the opportunity to point out that effects from
some of the lighter quarks (especially bottom) in the toponium system must be
treated more carefully than it is usually done by including the quarks only in
the number of flavours of a running coupling constant.

In sect. 4.2 we revisit the old calculation of Duncan \pcite{Dun} for the
one-loop vertex correction within our present formalism, avoiding some
approximations made in that early computation which allows us even some
statements on the term $O(\a_s^5m)$.

Sect. 5 is devoted to an exploration of possible further corrections of the
same numerical order of magnitude as $O(\a^4m)$. We list several candidates of
relevant QCD graphs and we show in cases of the most simple two loop graphs
that such corrections may well be relevant.  Beside these graphs from QCD,
also corrections from the weak interaction and QED may turn out to contribute
to this order, but the nonlocal Schwinger-Christ-Lee type graphs, peculiar to
the Coulomb gauge, are irrelevant to $O(\a^4m)$ as shown by an explicit
calculation. Among others we observe that the effect of the Z -Boson has been
underestimated so far since it gives rise to a singlet triplet splitting which
is equally important as the usual Breit interaction.

It should be stressed that our results for the energy shifts - below threshold
are - of course - in principle applicable to the bottonium system as well. In
fact a "rigorous" derivation of the corrections in that case has been
presented in the last ref. \pcite{scatt}, taking into account perturbatively
the gluon condensate as a description of the confinement effects. In view of
the large magnitude of that correction in the $b\bar{b}$ system and of the
also relatively large theoretical error of the latter we feel safer in the
$t\bar{t}$ system where those nonperturbative contributions are completely
negligible.

In the last section 7 we summarize our potential for $t\bar{t}$ in the
threshold region, including all terms which lead to numerical orders $O(m
a_s^4)$ in the level shifts, as discussed in our present paper. Still missing
pieces to that order are emphasized.


\section{BS-Perturbation of the BR Equation}
\plabel{Form}

A correct formulation of QCD in Coulomb gauge entails not only
Faddeev-Popov-ghost terms but also the inclusion of nonlocal interaction terms
\pcite{Schw}. Therefore, the full Lagrangian reads ($a$=1,...,8 for SU(3)):
\begin{equation} \plabel{Lag}
  {\cal L} = -\frac{1}{4} F_{\m\n}^{a} F^{a \m\n} + \sum_{j=1}^{f}
\bar{\Psi}_j (i \g D -m_j) \Psi_j + B^a (\6_j A_j^a)-\bar{\h}^a \6^i (\d_{ab}
\6_i + g f_{abc} A_i^c) \h^b + v_1 + v_2
\end{equation}
where the Lagrange multiplier $B^a$ guarantees the Coulomb gauge, and where
\begar
  D_{\m} &=& \6_{\m} - i g T^a A_{\m}^a, \\ F_{\m \n}^a &=& \6_{\m} A_{\n}^a
-\6_{\n} A_{\m}^a + g f_{abc} A_{\m}^b A_{\n}^c.
\ea
$v_1$ and $v_2$ are given in \pcite{Schw} and are discussed more explicitely
below.  The above Lagrangian will include all effects of the strong
interaction, but, as we will show, QED and weak corrections may also give a
contribution within the numerical order of our main interest (O($\a_s^4$)).

The BS equation in terms of Feynman amplitudes for $K$ and $S$ reads for a
bound state wave function $\chi$
\beg \plabel{allg}
   \chi_{ij}^{BS}(p;P) = -i S_{ii'}(\frac{P}{2}+p)S_{j'j}(-\frac{P}{2}+p) \int
\frac{d^4p'}{(2\pi)^4} K_{i'j',i''j''}(P,p,p')  \chi_{i''j''}^{BS}(p';P),
\ee
where $S$ is the exact fermion propagator, and $K$ is the sum of all two
fermion irreducible graphs. Furthermore, we have introduced relative momenta
$p$ and $p'$, a total momentum $P=p_1-p_2$, and we choose a frame where
$P=(P_0,\vec{0})=(2m+E,\vec{0})$.  The notation can be read off the pictorial
representation in fig.1.  $i,j$ are collective indices for spin ($\s,\r$) and
colour (noted $\a,\b$).

\subsection{Zero order equation}

It is well known that the dominant part in $K$ for weak binding ($\a \to 0$)
is the one-Coulomb gluon exchange which results in an ordinary Schr\"odinger
equation with static Coulomb potential. This result is even independent of the
chosen gauge in the ladder approximation by a simple scaling argument $p_0
\approx O(m\a^2), |\p| \approx O(m\a), P_0 \approx 2m-O(m\a^2)$ \pcite{Kum}:
\beg \plabel{Kc}
  K \to K_c \g^0 \otimes \g^0 := -\frac{4 \pi \a}{(\p-\p\,')^2} \g^0_{\s \s'}
\g^0_{\r' \r}
\ee
In equation \pref{Kc} we have already used the fact that only colour singlet
states can form bound states because the Coulomb potential is repulsive for
colour octets. The colour trace will always be understood to be already done,
leading to the definition
\beg
 \a \equiv \frac{4}{3} \frac{g^2}{4\pi} = \frac{4}{3} \a_s\plabel{alpha}
\ee
to be used in the following, in terms of the usual strong coupling constant
$\a_s$.  Because the above mentioned nonrelativistic limit of the BR equation
contains the projection operators $\l^\pm$, defined below, it is awkward to
calculate the so-called relativistic corrections in a straightforward way
within the framework of BS perturbation theory, starting from \pref{allg} with
\pref{Kc}.  Therefore, we use the BR equation \pcite{BR} instead of the
Schr\"odinger equation.

Moreover, for our present case, we need a generalization of \pcite{BR} for
unstable fermions, described by complex $m \to \tilde{m}=m-i \hat{\G_0}/2$
where $\hat{\G_0}/2$ represents a suitaby corrected (c.f. below) weak decay
rate of the free top quark. Since the real part of $\tilde{m}$ should still
determine the sum of polarizations in the numerator of the propagator, we take
by analogy to the stable case in the zero order approximation to
\pref{allg} the relativistic free propagator ($ E_p = \sqrt{\p\,^2+m^2},
\h =m \hat{\G_0}/2 E_p$)

\begar \plabel{Sg}
  S(\pm \frac{P_0}{2}+p_0) &\to& [(\pm \frac{P_0}{2}+p)\g - \tilde{m}]^{-1} =
\\ & & =  \frac{\L^+ \g_0}{\pm \frac{P_0}{2}+p_0-E_p+i \h} + \frac{\L^-
\g_0}{\pm \frac{P_0}{2}+p_0-E_p-i \h} + O(\frac{\G}{m})
\ea
with the relativistic projectors
\beg
 \L^{\pm}(\p) \equiv \frac{E_p \pm (\vec{\a} \p + \b m)}{2E_p}. \plabel{L}
\ee
If furthermore $\6 K_0/\6 p_0 = 0$ both sides of \pref{allg} may be
integrated with respect to $p_0$. On the r.h.s. the product of the two
propagators $S$ with \pref{Sg} yield four terms, with Cauchy poles determined
by $i\h$, two of which give no contribution.
Generically we obtain
\beg \plabel{D}
 \int \frac{d p_0}{2 \pi i} S \otimes S = \frac{\L^+ \g_0 \otimes \L^-
\g_0}{P_0-2E_p+2 i \h} + \frac{\L^- \g_0 \otimes \L^+ \g_0 }{-P_0+2E_p-2i\h}
\ee
so that $ K_{0} (\p, \p\,') \Phi_{0}$ with
\beg
  \Phi_{0} = \frac{1}{2 \pi} \int \chi_{0} dp_0
\ee
remains to be inserted at the places of "$\otimes$". As in \pcite{BR}
the kernel is now chosen as
\beg  \plabel{K0}
K_{0} \Phi_{0} = [\g_0 \L^+ \l^+ \L^+] \Phi_{0}  [\L^- \l^- \L^- \g_0 ]
\tilde{K}
\ee
so as to annihilate the second term in \pref{D}. In
\beg
(P_0 -2 E_p + 2 i \h) \Phi_0 = \int \frac{d^3p'}{(2\pi)^3}
\tilde{K}(\p,\p\,',P_0,i\h) \L^+ \l^+ \L^+ \Phi_{0} \tilde{\L}^- \l^-
\tilde{\L}^-
\ee
instead of $\L^-$ the projector $\tilde{\L}^- := \L^-(-\p)$ appears.  Thus
expanding $\Phi_{0} = \sum_{A,B=\pm} \L^{A} \tilde{\Phi}_{0}^{(AB)}
\tilde{\L}^{B}$ only $\Phi^{(+-)}_{0}$ is found to differ from zero and obeys
\beg  \plabel{LL}
(P_0 -2 E_p + 2 i \h) \L^+ \Phi_0^{(+-)} \tilde{\L}^- = \int
\frac{d^3p'}{(2\pi)^3} \tilde{K}(\p,\p\,',P_0,i\h) \L^+ \l^+
\Phi_{0}^{(+-)}(\p\,')  \l^- \tilde{\L}^-.
\ee
The nonrelativistic projection of \pref{LL} onto $\l^+ \otimes \l^-$ with
\begar
  \l^+ \L^{+} \l^+ &=& \m \l^+ \nonumber \\ \l^- \tilde{\L}^{-} \l^- &=& \m
\l^- \\ \m = \frac{E_p+m}{2 E_p}, \nonumber
\ea
and the introduction of appropriate factors in $\tilde{K}$ relatively to the
Coulomb kernel \pref{Kc}
\begar
  \tilde{K} &=& (\m \m' \n \n')^{-1} K_c \plabel{Ktil} \\
     \n ^2 &=& (P_0 + 2 E_p +2 i \h)/4 \nonumber
\ea
and in the wave function
\beg
 \Phi_{0}(\p) \propto \n \m \phi (\p)
\ee
also for the case of an unstable fermion lead to a Schr\"odinger equation
for the wave-functions $\Phi (\p)$ in momentum space ($\tilde{E}=
P_0-2m +2 i \h = E+2 i\h$):
\begar
[\frac{\p\,^2}{m} - \hat{E}_n]  \phi (\p) &=& \frac{\a}{(2 \pi)^2} \int
\frac{d^3 p'}{(\p -\p\,')^2} \phi(\p\,')  \plabel{Schr}
\ea
The eigenvalues for $\hat{E}_n=\tilde{E} + \frac{\tilde{E}^2}{4m}$ in
\pref{Schr} clearly occur at the real Bohr levels $E_n$, i.e.
\beg
 \tilde{E}_n = M_n^{0} - 2m = -\frac{m \a^2}{4 n^2} -\frac{m \a^4}{64 n^4} +
              O(\a^6)
\ee
In addition to the selection of the "large" components
by the choice \pref{K0}, obviously also the sign of $i\h$ in \pref{Ktil}
was crucial for the dependence of \pref{Schr} on the combination
$E+2 i \h$ alone. Still, for the bound-state argument at complex values
of the energy, the independence of $\h$ with respect to the momentum
$\p$ is essential. Going back to \pref{Sg} we observe that the
choice
\beg  \plabel{gh}
   \hat{\G}_0 = \frac{E_p}{m} \G_t
\ee
with $\G_t = const$ the (c.m.) decay rate of a single top quark, the
disturbing $\p$-dependence in $\h$ is cancelled. We note that \pref{gh} may be
interpreted as the width corrected by time dilatation.

The full BS-wave function (colour singlet, BS-normalized \pcite{Luri}) for the
BR-kernel can be obtained by going backwards to $\chi^{(+-)}_{0}:= \chi(p)$.
It appears as the real residue of the complex pole.  The wave function $(\o_n
= E_p -(2m+\tilde{E}_n)/2-i\e)$
\beg
\chi_n (p,\e)= \g_0 \bar{\chi}_n^*(p,-\e) \g_0 = i \frac{\L^+ S \tilde{\L}^-
}{(p_0^2 - \o_n^2) } \frac{\m(p)}{\n(p)} \frac{2 \o_n}{\sqrt{P_0}} \phi(\p).
\plabel{xBR}
\ee
is identical to the BR wave-function of a stable quark and belongs to the
spectrum of bound states $ P_n =M_n^0 - i \G_t $. The $i \e$ has been
introduced to determine the integration around that pole.

In eqs. \pref{xBR} $S$ is a constant $4\times4$ matrix which
represents the spin state of the particle-antiparticle system:
\beg \plabel{spin}
      S = \left\{ \begin{array}{l@{\quad:\quad}l}
                   \g_5 \l^- &  \mbox{singlet} \\
                \vec{a}_m \vec{\g} \l^- &  \mbox{triplet}.
                   \end{array}  \right.
\ee
$\phi$ is simply the normalized solution of the Schr\"odinger equation in
momentum space, depending on the usual quantum numbers $(n,l,m)$ \pcite{Wf},
$a_{\pm 1}, a_0$ in \pref{spin} describe the triplet states.
In the following it will
often be sufficient to use the nonrelativistic approximations of
eqs. \pref{xBR} ($\o_n \approx \tilde{\o}_n = (\p\,^2/m+E_n)/2 -i\e $)
\begar
  \chi(p)^{nr} &=& \frac{\sqrt{2} i \tilde{\o}_n}{p_0^2-\tilde{\o}_n^2}
\phi(\p)  S = -\g_0 \bar{\chi}(p)^{nr} \g_0 \plabel{xapp}
\ea
The net result of this subsection is that not only in the nonrelativistic
case \pcite{Khoze,Peskin}, but also for the BR-equation, by a suitable choice
of the kernel and of the wave function to zero order the exact solution
can be reduced to the one of the Schr\"odinger equation at the pole with
energy $E$ shifted to real values of $\tilde{E} = E + 2 i \h = E + i \G_t$,
where $\G_t$ is the decay rate of a single top quark in its rest system. Thus
the neccesary requirement is fulfilled to calculate a rigorous QCD potential
from its effects at such a pole. That this pole lies off the real axis is
unimportant. Another way to express the same fact is the following: Since the
only effect of the finite width $\G_t$ could be absorbed in a (complex) energy
alone, also for the BR-equation we could perform all calulations at real
energies $E$ (or $\tilde{E}= E+i \e$ if a singularity occurs on the real
axis), and continue analytically afterwards to $E=\tilde{E}-i \G_t$. Clearly
in this way the corresponding function in the upper $\tilde{E}$ plane is
obtained, because the Breit-Wigner pole (in our case ), of course, has a
complex conjugate counterpart at $\mbox{Im} \tilde{E} < 0$, seperated by a cut
along the real $\tilde{E}$ axis.

\subsection{Level-Shifts}

Having established that - as far as the systematic determination of the
QCD potential is concerned - we may now just consider perturbation
theory for "nonabelian positronium", perturbation theory for the BS equation
starts from the BR equation for the Green function $G_{BR}=G_0$ of the
scattering of two fermions \pcite{BR2}
\beg
 iG_0 = -D_0 + D_0 K_0 G_0,
\ee
which is exactly solvabel.
$D_0$ is the product of two zero order propagators, $K_0$ the corresponding
kernel.  The exact Green function may be represented as
\beg \plabel{Reihe}
 G =\sum_{l} \chi_{nl}^{BS} \frac{1}{P_0 - M_n} \bar{\chi}_{nl}^{BS} +
G_{reg}=G_0 \sum_{\n =0}^{\infty} (H G_0)^{\n} ,
\ee
where the corrections are contained in the insertions $H$.
Bound state poles $M_n$ contribute, of course, only for $P_0<2m$.
It is easy to show that
$H$ can be expressed by the full kernel $K$ and the full propagators $D$:
\beg \plabel{H}
 H = -K + K_0 +iD^{-1}-iD_0^{-1}.
\ee
Since the corrections to the external propagators contribute
only to $O(\a^5)$ \pcite{Male}, the perturbation kernel is essentially
the negative difference of the exact BS-kernel and of the
zero order approximation.

Expanding both sides of equation \pref{Reihe} in powers of $P_0-M_n^0$,
the mass shift is obtained \pcite{Lep77}:
\beg \plabel{dM}
 \D M = \< h_0 \> (1+\< h_1 \> ) + \< h_0 g_1 h_0 \> + O(h^3) .
\ee
Here the BS-expectation values are defined as e.g.
\begar
 \<h\> &\equiv& \int \frac{d^4p}{(2\pi)^4} \int \frac{d^4p'}{(2\pi)^4}
           \bar{\chi}_{ij}(p) h_{ii'jj'}(p,p') \chi_{i'j'}(p'), \plabel{erww}
\ea
We emphasize the four-dimensional p-integrations which correspond to the
generic case, rather than the usual three dimensional ones in a completely
nonrelativistic expansion. Of course, \pref{erww} reduces to an ordinary
"expectation value" involving $d^3p$ and $\Phi(\p)$, whenever $h$ does not
depend on $p_0$ and $p_0'$.

In \pref{dM} $h_i$ and $g_i$ represent the expansion coefficients of $H$ and
$G_0$, respectively, i.e.
\begar
 H&=& \sum_{n=0}^{\infty} h_n (P_0-M_n^0)^n \\
 G_0&=& \sum_{n=0}^{\infty} g_n (P_0-M_n^0)^{n-1}
\ea
Similar corrections for the wave functions \pcite{Lep77} are
irrelevant in our present work.

\setcounter{equation}{0}

\section{QCD Tree Diagrams}
\plabel{tree}

The contributions from the tree diagrams 2.a to 2.c
are well-known from the abelian case.  Fig. 2.a is peculiar
for the use of a different zero order equation than the Schr\"odinger
equation. It contains
the difference between the exact one Coulomb-gluon exchange and the
BR-Kernel ( \pref{K0} at $P_n-\G_t$). The exchange of one transverse gluon
is represented by graph 2.b. The annihilation graph fig. 2.c
with one gluon does not contribute in our nonabelian case.

For the sake of completeness and in order to illustrate the present formalism,
we exhibit first the results for the tree graphs as well.
The perturbation kernel for the Coulomb-gluon
exchange
\beg
  -iH_c := -ig^2 T^a_{\a \a'} T^a_{\b'\b} \g^0_{\s \s'} \g^0_{\r' \r}
\frac{1}{(\p-\p\,')^2},
\ee
is needed for the calculation of the energy shift induced by fig. 2.a using
Eqs. \pref{dM}, \pref{H}, \pref{erww}
and \pref{K0} to \pref{Ktil}. For the spin-singlet we have:
\begar
 \D M_{2.a} &=& \<H_c+K_{BR}\> = \nonumber \\
        &=&  \frac{\a^2}{16n^2}\<K_c\> + \frac{\pi \a}{P_0 m} \int
\frac{d^3p}{(2\pi)^3} \int \frac{d^3p'}{(2\pi)^3} [\phi^* \phi - 2 \phi^*
\frac{\p \p\,'}{(\p-\p\,')^2} \phi] + O(\a^6) = \nonumber \\ &=& m \a^4(
\frac{\d_{l0}}{8n^3} + \frac{1}{16n^4} - \frac{1}{16 n^3 (l+1/2)}) + O(\a^6) .
\plabel{dMc}
\ea

The contribution from the transverse gluon (fig. 2.b)
\beg
 -i H_{2.b} = i 4\pi \a \g^j_{\s\s'} \g^k_{\r'\r} \frac{1}{q^2}
(g_{ik}+\frac{q_j q_k}{\vec{q}\,^2}), \plabel{H1b}
\ee
with
\beg
      q\equiv p'-p,
\ee
gives rise to a spin singlet-triplet (magnetic hyperfine) splitting.  Because
of the $\g^j$ matrices, the $\l^{\pm}$ projectors from both wave functions
annihilate \pref{H1b}.  This means that two factors $\p \vec{\g}$, contained
in $\L^\pm$, are needed for a nonzero result which in turn gives rise to two
extra orders of $\a$. By this mechanism we arrive at the well known
contribution $O(\a^4)$ from this graph.  For the spin-singlet the mass shift
reads
\begar
\D M_{2.b,S=0}&=& \frac{2\pi \a}{m^2} [-|\Psi(0)|^2 + 2 \int
\frac{d^3p}{(2\pi)^3} \int \frac{d^3p'}{(2\pi)^3}  \phi^*(\p\,')
(\frac{(\p\q)(\p\,'\q)}{\q\,^4}-\frac{\p\p\,'}{\q\,^2}) \phi(\p) ]+ O(\a^6) =
\nonumber \\ &=&m\a^4(\frac{1}{8 n^4} -\frac{\d_{l0}}{8 n^3} -\frac{3}{16 n^3
(l+\2)})+ O(\a^6).  \plabel{dMT}
\ea
The evaluation of the singlet-triplet splitting requires some awkward
Dirac-algebra, but the final result may be brought into a quite transparent
form (where one recognizes this expression as the well known spin-spin and
spin-orbit interaction, adapted to the present problem, cf. e.g. \pcite{Land})
\begar
 \D M_{ortho,m} - \D M_{para} &=& \frac{2 \pi \a}{m^2} \int d^3p \int d^3p'
                      \phi^*(\p\,')(1+\frac{|\q \vec{a}_m|^2}{\q\,^2}-
                      3 \frac{(\p\,' \times \p)(\vec{a}^*_m \times \vec{a}_m)}
                      {\q\,^2} ) \phi(\p)= \nonumber\\
               &=&    \frac{2 \pi \a}{m^2}
        \< \frac{4}{3} \d(\vec{r}) + \frac{1}{4\pi}
        \frac{\vec{r}\,^2-3|\vec{r} \vec{a}_m|^2}{|\vec{r}|^5} -
\frac{3i(\vec{r} \times \p)(\vec{a}^*_m \times \vec{a}_m)}{4\pi r\,^3}
\>.\plabel{op}
\ea
Similarly as in the second line of \pref{op} all contributions from graphs 2
are summarized in the Breit interaction
\begar
 \CV_2 &=& -\frac{\p^4}{4 m^3} - \frac{\a \pi}{m^2} \d (\vec{r}) - \frac{\a
}{2 m^2 r} \left( \p\,^2 + \frac{\vec{r} (\vec{r} \p) \p} {r^2} \right) +
\nonumber\\ & & + \frac{3 \a }{2 m^2 r^3} \vec{L} \vec{S} +  \frac{\a }{2 m^2
r^3} \left( \frac{3  (\vec{r} \vec{S})^2} {r^2} - \vec{S}\,^2 \right) +
\frac{4 \pi \a }{3 m^2} \vec{S}\,^2 \d (\vec{r}) \plabel{v2}
\ea

\setcounter{equation}{0}

\section{One Loop Corrections}

\subsection{One Loop Vacuum Polarization}

In the case of positronium no massless particles are available to one loop
order in the vacuum polarization, so this effect is only of order of magnitude
$\a^5$.  In contrast to this, QCD contains massless gluons and light quark
flavours which may contribute significantly to the spectrum.  Usually such
terms are included in the "running coupling constant".  This quantity,
however, looses its (low order) gauge independence for non-vanishing masses.
Moreover, only {\it leading} logarithms in $|\q|$ are contained in that
approach. Thus beginning with two loops logarithmic terms are "lost" even for
massless loops.  In accordance with our systematic approach we thus evaluate
first only the loop in fig. 3.a by standard techniques and obtain
\begar
 \pi^{ab}_{3.a} &\equiv& 12 g^2  \d^{ab} \int \frac{d^Dr}{(2\pi)^D}
\frac{\q\,^2 - \frac{(\q \vec{r})^2}{\vec{r}^2}}{r^2 (\q-\vec{r})^2 }
\nonumber \\ &=& -\frac{i g^2 \d^{ab}}{ \pi^2} \q\,^2 [ \CD -\ln\q\,^2
+\frac{7}{3} - 2\ln2 -  \plabel{pi2er}\\ & & - \e(\frac{7}{3} - 2\ln2 -\g +\ln
4\pi) \ln \q\,^2 + \frac{\e}{2} \ln^2\q\,^2 + \e \cdot const + O(\e^2)]
\nonumber
\ea
with
\begar
     \CD &=& \frac{1}{\e}-\g+\ln 4\pi, \qquad \e = \frac{4-D}{2}. \plabel{Div}
\ea
We have included the term $\e\cdot f(q)$ in order to make it applicable in a
two loop calculation.  The graph 3.b contains $q_0$ terms which can be avoided
if we carry out the $p_0$ integration first (cf. \pref{erww} and \pref{dM}).
The result can be expanded in powers of $\p$ and $\a$ in order to show that
the effect of $q_0$ is of $O(\a^5)$.  With this simplification graph 3.b is
exactly calculable:
\beg \plabel{pi1er}
 \pi^{ab}_{3.b} = \frac{i g^2 \d^{ab}}{32 \pi^2}[ 10 \q\,^2 (\CD -\ln\q\,^2)
+16(7 - 8\ln2) \q\,^2 ] +O(\e)
\ee
Our renormalization prescription consists of a subtraction at the point
$q=(0,\vec{\m})$, where $\m$ has to be of the order $\a m$ to avoid large
logarithmic contributions from higher orders. This seems to be the natural
renormalization prescription for bound state problems, because also in the BS
expectation values \pref{erww}, the Bohr momentum $\a m$ together with $p_0
\approx O(\a^2 m)$ provides the dominant parts of the integrals
\pcite{KumW,KumWL}.

After renormalization, the contribution from the gluonic vacuum polarization
(with the colour trace already done) reads \pcite{Dun}
\begar \plabel{Kg}
-iH_g &=& -i \g_0 \otimes \g_0 \tilde{H_g}\\
\tilde{H_g} &=&-\frac{ 33 \a^2}{4 \q\,^2} \ln \frac{\q\,^2}{\vec{\m}^2}.
            \nonumber
\ea
The expectation value of this expression can be obtained by performing the
Fourier transformation into coordinate space, where the integrations can be
done analytically (see Appendix A).  The surprisingly simple result is
\beg \plabel{dMg}
 \D M_g = \<H_g\> = - m \a^3 \frac{11N}{16 \pi n^2 } [ \Psi_1(n+l+1)+\g + \ln
\frac{\m n}{\a m}] + O(\a^5)
\ee
where $\Psi_n$ is the n-th logarithmic derivation of the gamma function and
$\g$ denotes Euler's constant. The closed form of Eq. \pref{dMg} was not
obtained in previous calculations.

Of course, the contribution to the potential is the standard one:
\beg \plabel{vg}
\CV_g =-\frac{33 \a^2}{8 \pi r} ( \g + \ln \m r)
\ee

Now we turn to the fermion loops (fig. 3.c).  In the literature the lighter
quarks are usually taken as massless (and 'absorbed' in the number of flavours
appearing in the running coupling constant) or even ignored \pcite{Fein,Dun} ,
but we will show that they do contribute within the order of interest and,
furthermore, the explicit dependence on the masses of lighter quarks may be
important.  As pointed out already in the introduction, this is due to the
fact that the top quark is expected to lie above 170GeV \pcite{ALEPH,CDF}, and
therefore the bottom and charm quark can neither be taken as relatively
massless nor as relatively super-heavy as compared to the natural mass scale
$\a m$.

The finite part of the self energy in fig. 3.c is a well known quantity
\pcite{Land} for arbitary masses $m_j$:
\beg
 \Pi_F = -\frac{i g^2 \d^{ab}}{4\pi^2} \q\,^2 I(\q\,^2,m_j^2)
\ee
We approximate in the exact solution
\begar
 I(\q\,^2,m_j^2) &\equiv& \int_0^1 dx x(1-x) \ln [x(1-x)\q\,^2+m_j^2]
\nonumber \\ &=& \frac{1}{6} \ln m_j^2 -\frac{5}{18} + f(\r), \\ f(\r) &=&
\frac{2 \r}{3} + \frac{1}{6} (1-2\r) \sqrt{1+4\r} \ln \frac{\sqrt{4 \r
+1}+1}{\sqrt{4 \r +1}-1} \plabel{f},\\ \r &:=& \frac{m_j^2}{\q\,^2},
\ea
for later convenience $f(\r)$ by
\beg \plabel{fnaeh}
 f(\r) \approx \frac{1}{6} \ln (\frac{1}{\r} + e^{\frac{5}{3}}).
\ee
This agrees with the original $f(\r)$ better than 1\% within the whole
integration region.

It is easy to transform into coordinate space in order to obtain the
potential $\CV_F$, effectively produced by this fermionic vacuum polarization:
\begar
 \D M_F &=& \< \CV_F \>, \nonumber\\
  \CV_F(r) &=& -\frac{\a^2}{4\pi r} [\mbox{Ei}(-r m_j e^{\frac{5}{6}})
             -\frac{5}{6} + \2 \ln (\frac{\m^2}{m_j^2} + e^{\frac{5}{3}})].
             \plabel{HF}
\ea
We note that the energy shift produced by \pref{HF}
can be obtained in closed form  using  \pcite{Grad}
\begar \plabel{Eiint}
 \int_0^x e^{-\b x} \mbox{Ei}(-\a x) dx &=& -\frac{1}{\b} [ e^{-\b x}
\mbox{Ei}(-\a x)+ \ln (1+\frac{\b}{\a}) - \\ & & - \mbox{Ei}(-(\a+\b)x) ].
\nonumber
\ea
Thus a useful expression for the energy shift induced by fermionic vacuum
polarization with {\it arbitrary} masses $m_j$ reads
\begar
 \D M_F^j &=& -\frac{m\a^3}{8\pi n^2 } \lb \sum_{k=0}^{2n-2l-2} b_{nl}^k
\left[ \left( -\frac{d}{d\b} \right)^{2l+1+k} [-\frac{1}{\b} \ln (1+\b
\frac{\a m}{n m_j e^{\frac{5}{6}}})] \right]_{\b \to 1} - \right. \nonumber \\
& & \qquad \left. -\frac{5}{6} + \2 \ln (\frac{\m^2}{m_j^2} + e^{\frac{5}{3}})
\rb,  \plabel{KF}
\ea
with
\beg
  b_{nl}^k := \frac{(n-l-1)!}{k![(n+l)!]^3} \left(\frac{d}{d\r} \right)^k
[L_{n-l-1}^{2l+1}(\r)]^2 \Big|_{\r=0} .
\ee
For states up to $n=3$ we write this result more explicitly as
\beg
 \D M_{F,nl}^j = \frac{m\a^3}{8\pi n^2 } \lb A_{nl}(\frac{n m_j}{\a m})+ \ln
\frac{(\frac{\m^2}{m_j^2}+ e^{\frac{5}{3}})^{\2}}{e^{\frac{5}{6}} + \frac{\a
m}{n m_j} } \rb \plabel{KFex}
\ee
with $A_{nl}$ from Tab.2,\\[.5cm]
\begin{minipage}{16cm}
\centerline{\begin{tabular}{c|c|c}
 $n$ & $l$ & $A_{ln}(\frac{n m_j}{\a m})$ \\ \hline 1 & 0 & $a$ \\ 2 & 0 &
$a^3-\2 a^2 + a$\\ 2 & 1 & $\frac{1}{3}a^3+\2 a^2+a$\\ 3 & 0 & $2
a^5-\frac{7}{2}a^4+\frac{10}{3}a^3-a^2+a$\\ 3 & 1 & $a^5 - \frac{3}{4} a^4 +
\frac{1}{3} a^3 + \2 a^2 + a$\\ 3 & 2 & $\frac{1}{5}a^5 +\frac{1}{4}a^4
+\frac{1}{3}a^3+\2 a^2 + a$
\end{tabular}}
\vspace{.3cm}
\centerline{Tab. 2}
\vspace{.5cm}
\end{minipage}
using the shorthand
\beg
 a^{-1}:= 1+\frac{ n m_j e^{\frac{5}{6}} }{\a m}
\ee
Only for $m_j>>\a m$ this gives an $O(\a^5)$ Uehling term, modified by
off-shell subtraction, but for $m_j \to 0$ it becomes an $O(\a^3)$
contribution, which means that eq. \pref{KFex} interpolates numerically in a
range of two orders in $\a$. Therefore \pref{KFex} or, equivalently, \pref{HF}
with $m_j \ne 0$ must be definitely taken into account for quarks with $m_j
\approx \a m$ at $O(\a^4)$. Thus in our present case the finite mass of the
bottom quark $m_b \approx O(\a m)$ is important.

  \subsection{Vertex Corrections}

The one loop corrections to the vertex together with self-energy insertions
into the fermion lines (fig. 3.d) in the abelian case (positronium), are known
to provide corrections only of $O(\a^5)$. The reason for this is the
"classical" Ward identity which relates those contributions in such a way that
the sum of these terms vanishes at $|\q| \to 0 $. This Ward identity happens
to continue to hold even in the Coulomb gauge and even in the nonabelian case
\pcite{Fein}, but only for the vertex
corrections referring to the Coulomb component of the gauge field. However,
the presence of the gluon splitting graphs 3.e and 3.f produces a contribution
already to $O(\a^4)$ \pcite{Dun}. A simple dispersion theoretic argument
allows to understand this difference: In the abelian case the first graph 3.d
in the variable $|\q|$ for $q_0=0$ has a cut for Re$|\q|<2m$. Thus corrections
in $\q$, e.g. in the electron form factor $F_1$, for symmetry reasons must be
of order $\q\,^2$, because $F_1(\q\,^2)$ is regular at $\q \to 0$. This is no
longer the case with the mass zero intermediate state allowed in 3.e and 3.f.
The first - and to our knowledge only- computation of the nonabelian vertex
corrections in the sense of our present approach was performed in ref.
\pcite{Dun}. This work contains certain approximations which we want to avoid
in order to prepare the ground for future calculations even at the level
$O(\a^5)$. We thus make a systematic expansion and solve the remaining
integrals analytically which contain contributions of the order of interest.
The vertex correction of fig. 3.e after performing the colour trace becomes
\begar
 -iH_{3.e} &=& \frac{36 \pi^2 \a^2 q_i}{\q\,^2} \int \frac{d^4r}{(2\pi)^4}
\frac{1}{r^2 (\vec{r}-\q)^2} (-\d_{ki}+\frac{r_k r_i}{\vec{r}^2}) \times
\nonumber \\ & & \times [\g^k \frac{1}{\g p + \g r -m} \g^0 - \g^0 \frac{1}{\g
p' -\g r -m} \g^k] \nonumber \\ &=& \frac{36 \pi^2 \a^2 q_i}{\q\,^2} ( \g^k
v_{ki}(1,p)\g^0-\g^0 v_{ki}(-1,p') \g^k) \plabel{Vk}
\ea
where ( $\e = \pm 1$)
\begar
  v_{ki}(\e,p) &:=& \int \frac{d^4r}{(2\pi)^4}  \frac{1}{r^2 (\vec{r}-\q)^2}
(-\d_{ki}+\frac{r_k r_i}{\vec{r}^2}) \frac{\g p + \e \g r +m}{(p+\e r)^2 -m^2}
\plabel{vki}
\ea
After the $r_0$ integration it proves useful to proceed with the $p_0$
integrations (cf. \prefeq{erww} ), where in contrast to the author of ref.
\pcite{Dun}, who approximates already at this point, we took into account also
the pole arising from the denominator of Eq. \pref{vki}. This results in
\begar
 v_{ki}(\e,p) &=& -i \int \frac{d^3r}{(2\pi)^3}  \frac{-\d_{ki}+\frac{r_k
r_i}{\vec{r}^2}}{(\vec{r}-\q)^2} F(\p,\vec{r}) \\ F(\p,\vec{r}) &=&
\frac{[(r+E_{p+\e r})(m -\vec{\g}\p - \e \vec{\g} \vec{r}) +\g_0 E_{p+\e r}
\frac{P_0}{2}](r+E_{p+\e r}+\o)- \g_0 E_{p+\e r} \frac{P_0}{2} \o} {2 r
E_{p+\e r} (r+ E_{p+\e r})[ \frac{P_0^2}{4} - (r+ E_{\p+\e\vec{r}} +\o)^2]}.
\nonumber \\
\ea
The calculation of the $\g$ trace to lowest order requires the inclusion of
the $\vec{\g}\p$-terms in the wave functions \pref{xBR} from the projection
operators \pref{L}.  After performing this trace, we expand in terms of $\p$
and $m\a^2$ which enables us to combine both terms in eq. \pref{Vk}:
\beg
 H_{3.e} = \frac{36 \pi^2 \a^2}{\q\,^2} \int \frac{d^3r}{(2\pi)^3}
\frac{\q\,^2-\frac{(\q \vec{r})^2}{r^2}}{(\vec{r}-\q)^2} \frac{1}{2 r^2 E_r} +
\mbox{higher orders}   \plabel{Vnaeh}
\ee
Now Eq. \pref{Vnaeh} may be evaluated exactly in terms of dilogarithms and the
result has a cut for $|\q|<0$, but no pole at $|\q|=0$. It can be formally
expanded for $|\q|>0$ to $O(|\q|)$:
\beg
 H_{3.e} = \frac{9 \a^2}{|\q| m } (\frac{\pi^2}{8}-\frac{2 |\q|}{3}+ O(\q\,^2)
). \plabel{Va}
\ee
The first term  in this expansion has been obtained in \pcite{Dun}, the second
one is the expected contribution to $O(\a^5)$.  The BS - expectation value of
\pref{Va} becomes to $O(\a^4)$:
\beg
 \D M_{3.e} = \frac{9 m \a^4}{32 n^2} \frac{1}{n(l+\2)}. \plabel{nakorr}
\ee

{}From \pref{Va} we conclude that the effect of graph 3.e can be summarized in
the potential
\beg  \plabel{v3e}
 \CV_{3.e} = \frac{9 \a^2}{8 m r^2}
\ee

It remains to calculate the vertex corrections with two transverse gluons ,
graph 3.f. At first sight it seems that this graph would give a contribution
to order $\a^3$ because no $\p \vec{\g}$ terms are needed from the wave
functions. This, as we will see, is not true because the leading (constant)
term vanishes as a consequence of renormalization and accordingly graph 3.f
has been estimated to be of order $O(\a^5)$ in ref. \pcite{Dun}. Here we use
an approach which explicitely provides at least part of the exact result of
this contribution.  The vertex part of the graph 3.f reads:
\begar \plabel{Vb}
V_{3.f} &=& -\frac{3 g^3 T^b}{4} \int \frac{d^Dr}{(2\pi)^D} \frac{(q_0-2r_0)
\g^i (\g(p-r)+m) \g^k} {[(p-r)^2-m^2](r-q)^2 r^2} \times \\ & & \qquad \qquad
\times \left( \d_{ik}-\frac{(q_i-r_i)(q_k-r_k)}{(\q-\vec{r})^2} -\frac{r_i
r_k}{\vec{r}^2} + \frac{(q_i-r_i) \vec{r}(\q-\vec{r}) r_k} {\vec{r}^2
(\q-\vec{r})^2} \right) \nonumber
\ea
With the gamma-trace to relative $O(\a^2)$ and using Feynman parametrization,
the effective vertex part becomes
\begar
 V_{3.f}^{eff} &=& -\frac{3 i g^3  T^c}{16} \int \frac{d^{D-1}k}{(2\pi)^{D-1}}
\int_0^1 dx \int_0^{1-x} dy
(L^{-\frac{3}{2}}-3x^2m^2L^{-\frac{5}{2}})(1+\frac{[\vec{k}(\q-\vec{k})]^2}
{\vec{k}^2(\q-\vec{k})^2}), \plabel{vbna} \nonumber \\
\ea
where
\begar
              L &=& (y
q_0-xp_0)^2+\vec{k}^2+2x\p\vec{k}-2y\q\vec{k}-x(p^2-m^2)-y \q\,^2.
\ea
For our estimate it is sufficient to consider the part from the constant term
in the second bracket in \prefeq{vbna}.  The $\vec{k}$ integration can be done
easily, leaving us with a finite part
\begar
 V_{3.f,1}^{eff} &=&\frac{3 i g^3 T^c}{32 \pi^2} \int_0^1 dx \int_0^{1-x} dy
(\ln M+\frac{2 x^2 m^2}{M})
\ea
where
\begar
         M &=& x^2 m^2 - w,\\ w &=&y(1-y)q^2+x(1-x)(p^2-m^2)+2xypq.
\ea
This expression cannot be expanded in terms of $w$ because this would yield in
the next order a spurious linear divergence from the $q^2$ term which would
indicate an equally spurious $O(\a^4)$ contribution. Therefore, we expand in
terms of $(w+y(1-y)\q\,^2)$ and solve the leading part analytically in terms
of dilogarithms. Expanding the result in terms of $\q\,^2$, one has
\begar
     V^{eff}_{3.e} &=&\frac{i g^3 N T^c}{32 \pi^2}\{2-\frac{1}{4}[1+\ln(
\frac{\q\,^2}{m^2})]\frac{\q\,^2}{m^2} \}+ \mbox{other $O(\a^2)$}
\plabel{Veff}
\ea
Since \pref{Veff} does not contains a term $\propto |\q|$, the vertex
correction due to two transverse gluons does not contribute to $O(\a^4)$.

To one loop order also the box graph 3.g occurs in the correction to the BR
kernel. It possesses an exact counterpart in QED and is known to contribute
only to $O(\a^5)$ \pcite{Fult}.  Furthermore we have also investigated the
two-loop vertex-correction depicted in fig. 3.h. Of course this correction is
but one of several two loop vertex corrections. The renormalization must take
into account the whole set of these graphs. Nevertheless, it seems that after
proper renormalization they yield a contribution to $O(\a^5)$.

\setcounter{equation}{0}

\section{Other Corrections}

\subsection{Two Loop Vacuum Polarization}

As pointed out already in subsection 4.1, the usual renormalization group
arguments relying on massless quarks in the running coupling constant do not
consistently include the effect of 'realistic' quark masses in the toponium
system, when a systematic BS perturbation is attempted. However, in the one
loop case finite quark masses could yield terms of numerical order $O(\a^4)$,
therefore the same can be expected here, leading to corrections of $O(\a^5)$.
On the other hand, in a full calculation of effects of $O(\a^4)$ two loops
with gluons cannot be neglected. Although it is enough to consider the two
loop vacuum polarization for Coulomb gluons only, the computation of all those
graphs is beyond the scope of our present paper. We just want to indicate how
already the graphs 4.a-4.c yield contributions of this order which are {\it
non-leading} logarithms.  Performing the zero component integrations of
momenta results in (including a symmetry factor 1/2)
\begar
 -iH_{4.a} &=& -i \g_0 \otimes \g_0 \frac{9 g^6}{2 \q^4} (\Pi^{(1)}+\Pi^{(2)})
\ea with
\begar
\Pi^{(1)} &=& \int \frac{d^{D-1}k}{(2\pi)^{D-1}} \int
\frac{d^{D-1}p}{(2\pi)^{D-1}} \frac{1}{\p\,^2 |\vec{k}| |\q-\vec{k}-\p|},\\
\Pi^{(2)} &=& \int \frac{d^{D-1}k}{(2\pi)^{D-1}} \int
\frac{d^{D-1}p}{(2\pi)^{D-1}} \frac{[(\q-\vec{k}-\p)\vec{k}]^2}{\p\,^2
|\vec{k}|^3 |\q-\vec{k}-\p|^3}.
\ea
By using dimensional regularization, Feynman-parametrization and usual
integration formulas \pcite{Muta} we arrive at
\begar
 \Pi^{(1)} &=& \frac{\G(-\2+\e) \G^2(1-\e)}{\G^2(\2) (4\pi)^{\frac{3}{2}-\e}
               \G(2-2\e)} \int \frac{d^{D-1}p}{(2\pi)^{D-1}} \frac{1}{\p\,^2
               [(\q-\p)^2]^{-\2+\e}}= \plabel{Pi1} \\
           &=& \q\,^2 \frac{(4\pi)^{2\e} (\q\,^2)^{-2\e}}{\G^2(\2) (4\pi)^3}
\frac{\G^2(1-\e) \G(-1+2\e) \G(\2-\e)}{\G(\frac{5}{2}-3\e)} \nonumber
\ea
and
\begar
 \Pi^{(2)} &=& \frac{3 \G(\frac{3}{2}-\e) \G^2(1-\e)
\G(2\e)}{4(4\pi)^{3-2\e}(\frac{3}{2}-2\e)(-1+2\e) \G(\52) \G(\2) \G(\52-3\e)}
(\q\,^2)^{1-2\e}+ \nonumber\\ & & + \frac{3 \G(\2-\e) \G(2-\e)
(\q\,^2)^{1-2\e}}{2 (4\pi)^{3-2\e} \G(\frac{3}{2}) \G(\52)} \times
\plabel{Pi2} \\ & & \lb \frac{\G(2\e) \G(\frac{7}{2} - \e) \G(2-\e)}{(-1+2\e)
\G(\frac{3}{2}-\e) \G(\frac{9}{2} -3\e)} + \frac{\G(2\e)}{3-4\e} \left[
\frac{\G(1-\e)}{\G(\52-3\e)} -
\frac{4\G(2-\e)}{\G(\frac{7}{2}-3\e)}+\frac{4\G(3-\e)}{\G(\frac{9}{2}-3\e)}
\right] \right. \nonumber \\ & & \left.\qquad \qquad - \frac{2 \G(2\e)
\G(\52-\e) \G(2-\e)}{\G(\frac{3}{2}-\e) \G(\frac{9}{2}-3\e)}+ \frac{\G(2-\e)
\G(1+2\e)}{\G(\frac{9}{2}-3\e)} \rb. \nonumber
\ea
{}From eqs. \pref{Pi1} and \pref{Pi2} the finite, renormalized contribution to
the perturbation kernel may be extracted as
\beg
 H_{4.a} = \g_0 \otimes \g_0 \frac{81 \a^3}{8 \pi \q\,^2} \ln
\frac{\q\,^2}{\vec{\m}^2}.  \plabel{H3a}
\ee
Eq. \pref{H3a} differs from eq. \pref{Kg} by a simple factor proportional to
$\a$ and therefore results in the mass shift
\beg  \plabel{dM4}
 \D M_4 = \frac{81 m \a^4}{64 \pi^2 n^2 }(\Psi_1(n+l+1)+\g + \ln \frac{\m
n}{\a m}).
\ee
The contribution from fig. 4.b is similar. After performing the integrations
over the zero components we have
\begar
 \Pi_{4.b} &=& -\frac{i g^4 27 \d_{ab}}{2} \int \frac{d^{D-1}k}{(2\pi)^{D-1}}
\int \frac{d^{D-1}p}{(2\pi)^{D-1}} \frac{1}{(\q-\vec{k})^2 |\p| |\vec{k}|
(\q-\p)^2} \times  \\ & & \times \left[\q\,^2 -
\frac{(\q\vec{k})^2}{\vec{k}^2}- \frac{(\q\p)^2}{\p\,^2} +\frac{(\q\p)(\p
\vec{k})(\q\vec{k})}{\p\,^2 \vec{k}^2} \right] .  \nonumber
\ea
This expression can be written in terms of integrals already solved in the
course of the one loop calculation:
\begar
 \Pi_{4.b} &=& -\frac{i g^4 \q\,^2 3 \d_{ab}}{8 \pi^4} [ \CD^2 + 2 \CD
(\frac{7}{3}-2\ln 2) -2 \CD \ln \q\,^2  \plabel{Pi3b}\\ & & +2(\g-\ln 4\pi -
\frac{14}{3} +4 \ln 2) \ln\q\,^2 + 2 \ln^2 \q\,^2 + const+O(\e)]. \nonumber
\ea
Eq. \pref{Pi3b} contains overlapping divergences and two graphs like 3.a with
one or the other vertex replaced by a counterterm have to be added. After
that, only an additive infinity survives which is subtracted by our usual
renormalization prescription.  Graph 4.b has a net contribution which is
proportional to $\ln^2$:
\beg \plabel{Pserg}
 \Pi_{4.b}^{ren} = -i\frac{3 g^4 \q\,^2}{8\pi^4}\ln^2
\frac{\q\,^2}{\vec{\m}^2}.
\ee
The last two loop correction we are considering is the one in fig. 4.c, whose
contribution to the Coulomb gluon propagator can be written as
\begar
 \Pi_{4.c} &=& -54 ig^4 \int \frac{d^Dr}{(2\pi)^D} \int \frac{d^Dk}{(2\pi)^D}
(q-k)^r q^l \frac{1}{(\q-\vec{k})^2 (\q-\vec{k}-\vec{r})^2 r^2 k^2} \times
\nonumber \\ & &  \qquad \times \left( \d^{rl}-\frac{k^rk^l}{\vec{k}^2}
-\frac{r^rr^l}{\vec{r}\,^2}+\frac{r^r(\vec{r}\vec{k})k^l}{\vec{r}\,^2\vec{k}^2}
\right).  \plabel{Pi3c}
\ea
Eq. \pref{Pi3c} can be evaluated entirely in terms of gamma functions by a
somewhat lengthy calculation, but following the same steps as above.  The
analytic result is
\begar
 \Pi_{4.c} &=& -ig^4 \frac{27}{2(8\pi^2)^2} (\q\,^2)^{1-2\e} (4\pi)^{2\e}
\frac{\G{(\e)} \G{(2\e)} \G^2(2-\e) \G(\frac{1}{2}-\e) \G(\frac{1}{2}-2\e)}
{\G(\frac{5}{2}-2\e) \G(\frac{5}{2}-3\e) \G(1+\e)} (1-2\e)(1-4\e).\nonumber \\
\ea
Expanding in terms of $\e$ and properly renormalizing the result we finally
obtain
\beg
 \Pi_{4.c} = \frac{3 i g^4 }{16 \pi^4} (-\frac{4}{3}
\ln\frac{\q\,^2}{\vec{\m}^2}+\ln^2\frac{\q\,^2}{\vec{\m}^2})
\ee
In table 3 we collect the results for the self energy parts of fig. 4.a to
4.c, apart from a factor $-i\frac{3 g^4}{8 \pi^4} \q\,^ 2 $\\
\centerline{\begin{tabular}{c|c}
graph & $\Pi $ \\ \hline 4.a & $ \frac{3}{4}  \ln \frac{\q\,^2}{\vec{\m}^2}$\\
4.b & $ \ln^2 \frac{\q\,^2}{\vec{\m}^2} $ \\ 4.c & $
\frac{2}{3}\ln\frac{\q\,^2}{\vec{\m}^2} - \frac{1}{2}
\ln^2\frac{\q\,^2}{\vec{\m}^2} $
\end{tabular}}\\[.1cm]
\centerline{Tab. 3}\\[.3cm]
The full calculation of the gluon self energy to two loops seems to be very
involved in the Coulomb gauge.  However, already our partial results show the
importance of nonleading logarithms.

Two loop calculations of the vacuum polarization are even more difficult if
massive fermions are included. However, in view of the results from the one
loop calculation with massive fermions we may expect that nonvanishing masses
tend to decrease the importance of such terms in practice to something that
would be de facto numerically equivalent to $O(\a^5)$.  We try to circumvent
these problems, for the time being, by the following argument, which also
includes the three 'massless' quarks u,d,s.  Because of the Ward identity for
the Coulomb-vertex, it is clear from the theory of renormalization group that
the same corrections can be obtained by expanding the running coupling
constant with a two loop (gluons+u,d,s) input for the latter which provides
also the first 'nonleading' logarithmic contributions.  Our present
calculation in any case illustrates on the one hand the procedure to be
followed in a systematic BS perturbation theory. On the other hand, we believe
that especially the computation methods for the notoriously difficult Coulomb
gauge may be useful elsewhere.

The beta function to two loops is renormalization scheme independent for
massless quarks \pcite{Muta} and its two loop part has been calculated some
time ago \pcite{Jon}:
\begar
 \b(g) &=& -\b_0 g^3 - \b_1 g^5 - ...\\ \b_0 &=& \frac{1}{(4 \pi)^2}
(11-\frac{2}{3} n_f)\\ \b_1 &=& \frac{1}{(4 \pi)^4} (102-\frac{38}{3} n_f).
\ea
Here $n_f$ is the number of effectively massless flavours and $\b(g)$ is the
solution of
\begar
 \ln \frac{\sqrt{-q^2}}{\m} &=& \int_g^{\bar{g}} \frac{dg4}{\b(g)},
\ea
which reads up to two loops
\beg
  \ln \frac{\sqrt{-q^2}}{\m} = \frac{1}{2 \b_0} \left[
\frac{1}{\bar{g}^2}-\frac{1}{g^2} + \frac{\b_1}{\b_0} \ln \frac{\bar{g}^2
(\b_0+\b_1 g^2)}{g^2 (\b_0+\b_1 \bar{g}^ 2)} \right].
\ee
Considering this as an equation for $\bar{g}=g(\q\,^2)$  we 'undo' the
renormalization group improvement by expanding with 'small' $g^2 \propto \a$
(cf. eq. \pref{alpha} ):
\begar
 \a(\q\,^2)&=& \a \left\{ 1- \a \frac{33-2 n_f}{16 \pi}\ln
\frac{\q\,^2}{\m^2}+ \right.\\ & &  \qquad + \left. \frac{\a^2}{(16
\pi)^2}[(33-2 n_f)^2 \ln^2 \frac{\q\,^2}{\m^2}- 9(102-\frac{38}{3}n_f)\ln
\frac{\q\,^2}{\m^2}] \right\} \nonumber
\ea
Clearly the one loop term agrees with the calculation in subsect.  4.1 in the
limit $m_j \to 0$. That limit, however, is not appropriate here, as stated
above (cf. sect. 4.1), because we would loose in this way terms of numerical
$O(m\a^4)$.  For the computation of the rest we need the expectation value of
$(\ln^2 \frac{\q\,^2}{\m^2})/ \q\,^2$. This integral can be done analytically
(Appendix A) and the result is:
\begar
\<\frac{ \ln^2 \frac{\q\,^2}{\m^2}}{\q\,^2} \> &=& \frac{m \a}{2 \pi n^2}
       \{ \frac{\pi^2}{12} + \Psi_2(n+l+1)+s_{nl}+[\Psi_1(n+l+1)+\g+ \ln
\frac{\m n}{\a m}]^2 \}  \plabel{eln2}
\ea
with
\begar
s_{nl} &=& \frac{2 (n-l-1)!}{(n+l)!} \sum_{k=0}^{n-l-2} \frac{(2l+1+k)!} {k!
(n-l-1-k)^2}. \nonumber
\ea
With eq. \pref{eln2} we obtain for the mass shift, induced by the leading logs
of the two loop vacuum polarization of the Coulomb gluon a contribution:
\begar
 \D M_{2loop} &=& -\frac{m \a^4}{128 \pi^2 n^2} \left\{ 27^2[ \frac{\pi^2}{12}
+ \Psi_2(n+l+1)+s_{nl}+(\Psi_1(n+l+1)+\g+ \ln \frac{\m n}{\a m})^2] \right. +
\nonumber \\ & & \qquad \qquad \qquad+ \left.288 (\Psi_1(n+l+1) + \g +\ln
\frac{\m n}{\a m}) \right\}.  \plabel{dM2l}
\ea
In this expression we have set $n_f=3$ as dictated by the number of
sufficiently light quarks.  Whether eq. \pref{dM2l} really represents the full
two loop quark- gluon vacuum polarization, numerically consistent with other
terms $O(\a^4)$, must still be checked in a calculation of the Coulomb gluon's
self-energy to two loop order in the Coulomb gauge, i.e. going beyond the
sample calculation here.  We, nevertheless, indicate the corresponding
potential
\begar
 \CV_{2loop} =- 2 \frac{\a^3}{(16 \pi)^2 r} \left\{
(33-2n_f)^2[\frac{\pi^2}{6}+ 2(\g+\ln \m r)^2] + 9 (102-\frac{38}{3}
n_f)(\g+\ln \m r) \right\} \plabel{v2l}
\ea

\subsection{QCD 2-Loop Box Graphs}

It would be incorrect to extrapolate from the QED case the absence of
corrections to $O(\a^4)$, other than the abelian tree graphs because gluon
splitting allows new types of graphs.  Our first example of a QCD box graph is
fig. 5.a. Between the nonrelativistic projectors $\l^{\pm}$ of the wave
functions the perturbation kernel from this graph can be written effectively
as
\begar
 -i H_{5.a} &=&-12 i g^6 \int \frac{d^4t}{(2\pi)^4} \frac{d^4k}{(2\pi)^4}
\frac{(k_0+p_1^0+m)(q^0-t^0+p_2^0-m)}{[(p_1-k)^2-m^2][(p_2-t)^2-m^2]
\vec{k}\,^2 (\q-\vec{k})^2 \vec{t}\,^2 (\q-\vec{t})^2} \times  \nonumber \\ &
& \quad \times \frac{1}{(t-k)^2} \left(-(\q-\vec{k})\vec{k}+
\frac{[(\vec{t}-\vec{k})(\q-\vec{k})][\vec{k} (\vec{t}-\vec{k})]}
{(\vec{t}-\vec{k})^2} \right). \plabel{Hgraph}
\ea
After performing the integrations over the zero components $t^0$ and $k^0$ the
external momenta can be set to $(m,\vec{0})$. This is justified a posteriori
by the finiteness of the remaining terms.  The resulting expression will thus
only depend on $\q$ and $m$. The leading contribution should be
\beg
 H_{5.a} = c_{5.a} \frac{\a^3}{ |\q|^2}, \plabel{H5a}
\ee
but a really reliable estimate or even a calculation of the coefficient
$c_{5.a}$ is not available yet.  Supplementing the usual three powers of $\a$
from the wave functions for the computation of energy levels, we see that the
graph 5.a indeed would give a contribution $O(\a^4)$, expressible as yet
another correction $ \a \to \a (1 + c_{5.a} \a^2) $ in $\CV$ which certainly
is not included e.g. in the running coupling constant. The qualitative result
\pref{H5a} had been noted already in
\pcite{Dun}, \pcite{Fisch} and \pcite{Fein}.
It should be noted, however, that also e.g. the graph 5.b may yield a
contribution of the same structure.  A similar graph with crossed Coulomb
lines (fig. 5.c) is irrelevant because its group theoretic factor vanishes.

The 'X` graph in fig. 5.d can be checked more easily for possible new
contributions.  As in the calculation of fig. 5.a it can be simplified  to
give
\begar
 -i H_{5.d} &=& 3 i g^6 \int \frac{d^4t}{(2\pi)^4} \frac{d^4k}{(2\pi)^4}
\frac{(k_0+p_1^0+m)(t^0+p_2^0-m)}{[(k+p_1)^2-m^2][(t+p_2)^2-m^2] \vec{k}\,^2
(\q-\vec{k})^2} \times  \nonumber \\ & & \quad \times \frac{1}{t^2 (q-t)^2}
\left( 1+ \frac{[\vec{t}(\q-\vec{t})]^2} {\vec{t}\,^2 (\q-\vec{t})^2} \right).
\plabel{Xgraph}
\ea
The integration over $k$ yields a divergence $1/|\q|$ if $\q$ and $\p$ tend to
zero. Contributions within the order of interest may only result from possible
poles after the $t$-integration. For simplicity we consider the part of eq.
\pref{Xgraph} from the factor one in the second line:
\begar
 I_t &\equiv& \int \frac{d^4t}{(2\pi)^4} \frac{t^0+p_2^0+m}{[(t+p_2)^2-m^2]
t^2 (q-t)^2} = \nonumber\\ &=& \frac{-i}{(4\pi)^2} \int_0^1dx \int_0^{1-x} dy
\frac{y q^0 -x p_2^0+p^0+m} {(y q-x p)^2-y q^2 -x(p^2-m^2)} \approx
\nonumber\\ &\approx&  \frac{im}{(4\pi)^2} \int_0^1dx \int_0^{1-x} dy \frac{x}
{x^2 m^2+ y(1-x-y)\q\,^2} + O(\a)= \nonumber\\ &=& \frac{-i}{(4\pi)^2m} \ln
\frac{|\q|}{m} + O(\a).
\ea
Thus the part of graph 5.d, specified above, has a leading term from $$
H_{5.d,1} = \frac{3 g^6}{32(4\pi)^2 |\q| m} \ln \frac{|\q|}{m} $$ as $\q \to
0$ and therefore contributes to $O(\a^5 \ln \a)$. The second part of graph 5.c
gives a similar result with a different numerical factor.

We conclude that - opposite to the QED case \pcite{Fult}- box graphs may be
important to $O(\a^4)$. They could, in fact, be responsible for changes of the
zero order coupling of the Coulomb gluon.  Substantially different values of
that coupling seem to be required in phenomenological fits of lighter
quarkonia.

As illustrated by the explicit calculations above, to $O(\a^5)$ beside abelian
QED type corrections \cite{Male,Fult}, a host of further non-abelian
contributions can be foreseen.

\subsection{QED Correction}

As a rule, the consideration of electromagnetic effects in QCD calculations is
not neccessary, but at high energies the strong coupling decreases, and in the
case of toponium we expect $\a_s^2$ to be of the same order as $\a_{QED}$.  We
may obtain this contribution by simply solving the BR equation for the sum of
a QED and a QCD Coulomb exchange results in the energy levels
\begar
 P_0 =M_n^0 &=& 2m \sqrt{1-\frac{(\a+\a_{QED} Q^2)^2}{4n^2}} \approx  \\
&\approx& 2m - m \frac{\a^2}{4n^2} - \frac{m \a \a_{QED} Q^2}{2 n^2} - m
\frac{\a^4}{64n^4}  -\frac{m \a_{QED}^2 Q^4}{4 n^2} + O(\a^6), \nonumber
\ea
where $Q$ is the electric charge of the heavy quark, i.e. $2/3$ for toponium.
Clearly even the 'leading' third term can only be separated from the effect of
the second one to the extent that $\a(\m)$ and $\a_{QED}(\m)$ can be studied
separately with sufficient precision.

\subsection{Weak Corrections}

While also weak interactions usually can be neglected in QCD calculations,
this is not true in the high energy region, because the weak coupling scales
like $\sqrt{G_F m^2}$, becoming comparable to the strong coupling if the
fermion mass $m$ is large. Even bound states through Higgs exchange are
conceivable \pcite{Jain}. Therefore, we have to consider weak corrections and
especially the exchange of a single Higgs or Z particle, assuming for
simplicity the standard model with minimal Higgs sector.

The Higgs boson gives rise to the kernel
\beg
 -i H_{Higgs} = -i \sqrt{2} G_F m^2 \frac{1}{q^2-m_H^2}\approx i \sqrt{2} G_F
m^2 \frac{1}{\q\,^2+m_H^2}, \plabel{HH}
\ee
in an obvious notation.

Since we do not know the ratio $\a m/m_H$, which would allow some
approximations if that ratio is small, we calculate explicitly the level
shifts by transforming into coordinate space. As in Appendix A, we express the
Laguerre polynomials in terms of differentiations of the generating function,
do the integration and perform the differentiation afterwards to obtain
\beg
\D M_{Higgs} = -m \frac{G_F m^2 \a}{4 \sqrt{2} \pi} I_{nl}(\frac{\a m}{n m_H}),
\plabel{dMH}
\ee
valid for arbitrary levels and Higgs boson masses, with
\begar \plabel{inl}
 I_{nl}(a_n) &\equiv& \frac{a_n^{2l+2}}{n^2 (1+a_n)^{2n}} \sum_{k=0}^{n-l-1}
{n+l+k \choose k} {n-l-1 \choose k} (a_n^2-1)^{n-l-1-k}.
\ea
As an illustration some explicit results for the lowest levels are given in
tab.4.
\renewcommand{\arraystretch}{1.4}\\[.5cm]
\centerline{\begin{tabular}{c|cccccc}
 $n$ & 1 & 2 & 2 & 3 & 3 & 3 \\ \hline $l$ & 0 & 0 & 1 & 0 & 1 & 2 \\ \hline
$I_{nl}(a_n)$ & $\frac{a_1^2}{(1+a_1)^2}$ &
$\frac{a_2^2(2+a_2^2)}{4(1+a_2)^4}$ & $\frac{a_2^4}{4(1+a_2)^4}$          &
$\frac{a_3^2(3+6a_3^2+a_3^4)}{9(1+a_3)^6}$ &
$\frac{a_3^4(4+a_3^2)}{9(1+a_3)^6}$ & $\frac{a_3^6}{9(1+a_3)^6}$
\end{tabular}}

\vspace*{0.3cm}
\centerline{Tab. 4}
\vspace{.5cm}
It is evident that Eq. \pref{dMH} will give a contribution of order $G_F m^2
\a^3$ if the Higgs mass is comparable to the mass of the heavy quark and
should therefore be taken into account in a consistent treatment of heavy
quarkonia spectra and the related potential to numerical order $O(\a^4)$.  We
will thus consider also a corresponding term
\begar
 \CV_{Higgs} = - \sqrt{2} G_F m^2 \frac{e^{-m_H r}}{4 \pi r}
\ea
as a correction in our potential.

Next we consider the contributions of the neutral current, the single
Z-exchange and Z-annihilation. For the Z-exchange we obtain
\begar
 H_Z &=& -\sqrt{2} G_F m_Z^2 [\g^{\m}(v_f-a_f \g_5)]_{\s\s'}
\frac{g_{\m\n}-\frac{q_{\m}q_{\n}} {m_Z^2}}{q^2-m_Z^2}[\g^{\n}(v_f-a_f
\g_5)]_{\r'\r} \plabel{HZ}
\ea
with
\begar
 v_f &=& T^f_3-2 Q_f \sin^2 \Theta_w, \\ a_f &=& T^f_3,
\ea
where $T_3^f$ is the eigenvalue of the diagonal SU(2) generator for the
fermion $f$. If $f$ is the top quark then $T^f_3=1/2$.  Because we cannot
expect $\a m$ to be much smaller that $m_Z$ we use the exact expectation value
of the Yukawa potential \pref{inl}.  $q_0^2$ may be dropped in the
Z-propagator since this provides at most a correction $O(\a)$:

\begar
 \D M_Z &=& -\sqrt{2} G_F m_Z^2 \< \frac{a_f^2(3-2 \vec{S}\,^2)(1+
\frac{\q\,^2}{3 m_Z^2}) -v_f^2}{q\,^2-m_Z^2} \> = \\ &=& m \frac{G_F m_Z^2
\a}{2 \sqrt{2} \pi} [a_f^2(1-\frac{2}{3} \vec{S}\,^2) ( I_{nl}(\frac{\a m}{n
m_Z}) +\frac{(\a m)^2}{2 m_Z^2 n^3} \d_{l0} ) -\frac{v_f^2}{2} I_{nl}(\frac{\a
m}{n m_Z} ) ] \nonumber
\ea
$\vec{S}\,^2=S(S+1)$ is total spin of the quark-antiquark system. Therefore
this expression gives rise to a singlet triplet splitting within the order of
interest.

Z-annihilation may also yield a sizeable effect.  The corresponding energy
shift is easily evaluated
\begar
 \D M_{S=0} &=& \frac{3 G_F m^2 a_f^2}{2 \sqrt{2} \pi} \frac{m \a^3} {n^3}
\d_{l0} \\ \D M_{S=1} &=& -\frac{v_f^2}{a_f^2}\frac{m_Z^2}{4m^2-m_Z^2} \D
M_{S=0} \approx 10^{-2} \D  M_{S=0}
\ea
and also produces a singlet-triplet splitting.  It should be noted that the
two last contributions yield  singlet triplet splittings which are as
important as the usual Breit interaction \pref{v2}. The corresponding
contribution to $\CV$ is
\begar
 \CV_{Z} &=& \sqrt{2} G_F m_Z^2 a_f^2 \Big\{ \frac{e^{-m_Z r}}{2 \pi r}
\big[1-\frac{v_f^2}{2 a_f^2} - (\vec{S}\,^2 - 3
\frac{(\vec{S}\vec{r})^2}{r^2}) (\frac{1}{m_Z r} + \frac{1}{m_Z^2 r^2})-
(\vec{S}\,^2 - \frac{(\vec{S}\vec{r})^2}{r^2})\big]  + \nonumber     \\ & &
\qquad \qquad \qquad + \frac{\d(\vec{r})}{m_Z^2} (7- \frac{11}{3} \vec{S}\,^2
) \Big\} \plabel{vz}
\ea

\subsection{Schwinger-Christ-Lee Terms}

As mentioned in section \pref{Form}, nonlocal interactions have to be added to
the Lagrangian in Coulomb gauge.  We are not aware of any previous attempt to
look whether these terms give contributions to bound state problems or to some
effective potential.

By analogy to the second ref. \pcite{Schw} we calculate the $v_1$ term to
$O(g^4)$
\begar
 v_1 &=& -g^4 \frac{9}{16} \int d^3r d^3r' d^3r'' A_i^c(\vec{r}\,')
K_{ij}(\vec{r}-\vec{r}\,') K_{jk}(\vec{r}-\vec{r}\,'') A_k^c(\vec{r}\,'')\\
K_{ij}(\r) &=& \frac{1}{4 \pi |\r|} \left[ \frac{\d_{ij}}{3}
\d(\vec{\r})-\frac{1}{4 \pi |\vec{\r}|^5} (3 \r_i \r_j - \vec{\r}\,^2 \d_{ij})
\right]. \nonumber
\ea
This corrects the gluon propagator by
\beg
 \d G_{\m\n}^{ab}(x_1,x_2) = -\frac{1}{Z[0]} \frac{\d^2}{\d J_{\m}^a(x_1) \d
J_{\n}^b(x_2)} \frac{9 i g^4 }{16} \int d^4x \int d^3r d^3r' d^3r''
\frac{\d}{\d J_i^c} K_{ij} K_{jk} \frac{\d}{\d J_k^c} Z_0[J].
\ee
In momentum space $\d G$ can be calculated by using dimensional regularization
to give
\beg
 \d G_{mn}^{ab} (q,q') = (2\pi)^4 i \d^{ab} \d(q-q') \frac{9 g^4 }{8^5}
\frac{\q^2}{q^2} \frac{1}{q^2}(-\d_{mn}+\frac{q_m q_n}{\q^2}),
\ee
which means that we have a mass shift with the same structure as the one
transverse gluon exchange (cf. sect. \pref{tree}), further suppressed by two
more orders in $\a$.

Since the second term $v_2$ also represents a correction to the propagator of
the transverse gluon, it can be estimated by the same method to contribute
only in higher orders of $\a$ as well.  We thus find that both terms can be
neglected even including terms $O(\a^5)$.

\setcounter{equation}{0}

\section{Conclusions}

Both, the large mass \underline{and} the large width of the top quark provide
a new field for rigorous QCD perturbation theory: In contrast to the lighter
quarkonia, (unstable) toponium is a weakly bound system, to be treated by
Bethe-Salpeter methods in a systematic manner. The large width even further
reduces the effects of confinement. As shown first by Fadin and Khoze
\pcite{Khoze} $\G$ can be included in the (weakly bound) Green function at the
threshold in a straightforward manner by analytic continuation of the total
energy into the complex plane.

In our present paper we first show that the same trick may be also applied to
a different zero order equation, the BR-equation.  On the basis of this result
we describe how to obtain the proper potential $\CV$ for such a Green
function, rigorously derived from QCD in a perturbative sense. Although a
fully 4-dimensional formalism is used, which especially also allows the
inclusion of off-shell effects, our result allows an interpretation as a
correction to the Coulomb potential:
\begar
 \CV &=& -\frac{\p^4}{4 m^3} - \frac{\a \pi}{m^2} \d (\vec{r}) - \frac{\a }{2
m^2 r} \left( \p\,^2 + \frac{\vec{r} (\vec{r} \p) \p} {r^2} \right) +
\nonumber\\ & & + \frac{3 \a }{2 m^2 r^3} \vec{L} \vec{S} +  \frac{\a }{2 m^2
r^3} \left( \frac{3  (\vec{r} \vec{S})^2} {r^2} - \vec{S}\,^2 \right) +
\frac{4 \pi \a }{3 m^2} \vec{S}\,^2 \d (\vec{r}) \nonumber \\ & & - \frac{33
\a^2}{8 \pi r} ( \g + \ln \m r) +\frac{\a^2}{4\pi r} \sum_{j=1}^{5}
[\mbox{Ei}(-r m_j e^{\frac{5}{6}}) -\frac{5}{6} + \2 \ln (\frac{\m^2}{m_j^2} +
e^{\frac{5}{3}})]  \plabel{result}\\ & & + \frac{9 \a^2}{8 m r^2}  +  c_{5.a}
\frac{4\pi \a^3}{r} -  \frac{2 \a^3}{(16 \pi)^2 r} \left\{
27^2[\frac{\pi^2}{6}+ 2(\g+\ln \m r)^2] + 576(\g+\ln \m r)\right\}   \nonumber
\\ & & - \frac{8}{9} \frac{4\pi \a_{QED}(\m) \a}{r}- \sqrt{2} G_F m^2
\frac{e^{-m_H r}}{4 \pi r} + \sqrt{2} G_F m_Z^2 a_f^2
\frac{\d(\vec{r})}{m_Z^2} (7- \frac{11}{3} \vec{S}\,^2 )  \nonumber \\ & & +
\sqrt{2} G_F m_Z^2 a_f^2  \frac{e^{-m_Z r}}{2 \pi r}
\big[1-\frac{v_f^2}{2a_f^2} - (\vec{S}\,^2 - 3 \frac{(\vec{S}\vec{r})^2}{r^2})
(\frac{1}{m_Z r} + \frac{1}{m_Z^2 r^2})- (\vec{S}\,^2 -
\frac{(\vec{S}\vec{r})^2}{r^2})\big] \nonumber
\ea
It has been obtained by checking contributions up to numerical order $O(\a^4)$
to real energy shifts, calculated independently from the weak decay $t \to b +
W$. The first and the second line simply comprise the analogue of the
relativistic abelian corrections.  The third line corresponds to the one-loop
vacuum polarization effects for the gluon and from massive flavours. $m_j=0$
may be used for u,d,s-quarks, but not for c and b. The first term in the
fourth line results from the nonabelian vertex correction.  Among the terms
yielding $O(\a^4)$ in the other entries of the fourth line not all
coefficients are known at present : The "H"-graphs (5.a+5.b) and the two loop
terms involving a heavy flavour loop (bottom) still await an evaluation.  On
the other hand, we show that certain electroweak effects (QED,Higgs,Z) are
numerically important to $O(\a^4)$. Their respective contributions are listed
in the last two lines of
\pref{result}.
Comparing \pref{result} with QCD-potentials proposed previously
\pcite{scatt,Buchm}, the most important difference is that the
coupling $\a_s = 3/4 \a$ does not contain a "running" part.  Instead, terms
which usually (with zero mass flavours) are included in that running constant
have been written explicitely so that orders of the coupling constant are {\it
not} mixed.  This we believe to be important for an approach which guarantees
gauge-independence up to certain order. Of course, all calculations were
performed in the Coulomb gauge and $\CV$ should also be used in further
calculations {\it only} in that gauge.

Within a rigorous field-theoretic philosophy it would also be incorrect to
add, say, a linear term to \pref{result} in order to describe confinement. At
best \pref{result} could be supplemented by a piece $\propto \< G^2 \> r^3$
which mimics the tail of confinement effects by gluon condensate
\pcite{Leut,KumWL}.

In the derivation of our potential we have not only used the level shifts, but
also have described in much detail new closed forms for such shifts etc. The
reason for that has been that on the one hand we hope to have given new useful
methods to be applicable also for nonabelian $O(\a^5)$-effects. On the other
hand certain computations of level shifts may be useful in conjunction with
semi-phenomenological approaches to the lighter quarkonia.

\vspace{.5cm}
This work is supported in part by the Austrian Science Foundation (FWF) in
project P10063-PHY within the framework of the EEC- Program "Human Capital and
Mobility", Network "Physics at High Energy Colliders", contract CHRX-CT93-0357
(DG 12 COMA).

 \begin{appendix} \setcounter{equation}{0} 

\section{Expectation Values}

In sect. 4 and 6 we needed the expectation values of logarithmic potentials
between Schr\"odinger wave functions. They can be obtained by
\beg
 \< \frac{\ln^nr}{r} \> = \frac{d^n}{d\l^n} \< r^{\l-1} \> \big|_{\l=0},
\ee
if the expectation value $\< r^{\l-1} \> $ is known analytically.  In the
latter the representation
\beg
 L_{n-l-1}^{2l+1}(\r) = \lim_{z\to 0} \frac{1}{(n-l-1)!}
\frac{d^{n-l-1}}{dz^{n-l-1}} (1-z)^{-2l-2} e^{\r \frac{z}{z-1}}
\ee
of the Laguerre polynomials may be used. This allows an easy evaluation of the
integrations and the remaining differentiations can be done with some care
afterwards:
\beg
 \< r^{\l-1} \> = \frac{(\a m)^{1-\l }}{2 n^{2-\l } } \frac{(n-l-1)!}{(n+l)!}
\G(2l+2+\l) \sum_{k=0}^{n-l-1} {\l \choose   n-l-1-k }^2 {-2l-2-\l \choose k }
(-1)^k
\ee
Using now the Fourier transformations \pcite{Gelf}
\begar
 F[ \frac{\ln\frac{\q\,^2}{\m^2}}{\q\,^2}] &=& -\frac{\g+\ln \m r}{2\pi r}\\
F[ \frac{\ln^2\frac{\q\,^2}{\m^2}}{\q\,^2}] &=& \frac{1}{2\pi r}
[\frac{\pi^2}{6} + 2(\g+\ln \m r)^2]
\ea
one arrives immediately at eq.\pref{dMg} and \pref{eln2}, respectively.

 \end{appendix}

\u{Figure Captions}
\begin{itemize}
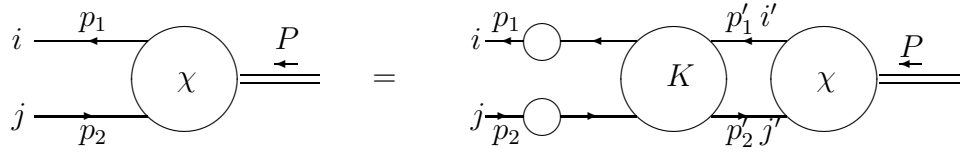
\item [Fig.1:] BS-equation for bound states
\item [Fig.2:] Tree graphs (broken lines represent Coulomb gluons, curly lines
depict transverse gluons, wavy lines represent a general gluon and solid lines
stand for fermions).
\item [Fig.3:] One loop graphs and vertex corrections
\item [Fig.4:] Two loop vacuum polarization
\item [Fig.5:] Two loop box graphs
\end{itemize}
\newpage

\unitlength1.0cm
\begin{picture}(18,5)
  \put(.5,3.5){\line(1,0){1.5}}
\put(3.24,3.05){\line(1,0){1.05}} \put(.5,2.5){\line(1,0){1.5}}
\put(2.5,3){\circle{1.5}}\put(3.24,2.95){\line(1,0){1.05}}
\put(1.5,3.5){\vector(-1,0){0.3}}    \put(2.4,2.9){$\chi$}
\put(1,2.5){\vector(1,0){0.3}} \put(1.1,3.7){$p_1$}
\put(3.7,3.4){$P$} \put(1.1,2.2){$p_2$}
\put(4,3.2){\vector(-1,0){0.3}} \put(0.2,3.4){$i$} \put(0.2,2.4){$j$}
\put(5,2.9){=}

  \put(6.5,3.5){\line(1,0){.5}} \put(7.25,3.5){\circle{.5}}
\put(7.5,3.5){\line(1,0){1}} \put(6.5,2.5){\line(1,0){.5}}
\put(7.25,2.5){\circle{.5}} \put(7.5,2.5){\line(1,0){1}}
\put(7,3.5){\vector(-1,0){0.3}}
\put(8.2,3.5){\vector(-1,0){0.3}} \put(6.5,2.5){\vector(1,0){0.3}}
\put(7.7,2.5){\vector(1,0){0.3}} \put(9.5,3.5){\line(1,0){1}}
\put(9,3){\circle{1.5}} \put(9.5,2.5){\line(1,0){1}} \put(8.9,2.9){$K$}
\put(11.74,3.05){\line(1,0){1.05}} \put(11,3){\circle{1.5}}
\put(11.74,2.95){\line(1,0){1.05}} \put(10.9,2.9){$\chi$}
\put(10.2,3.5){\vector(-1,0){0.3}}
\put(12.3,3.2){\vector(-1,0){0.3}} \put(9.7,2.5){\vector(1,0){0.3}}
\put(12,3.3){$P$} \put(6.6,3.7){$p_1$}
\put(9.7,3.7){$p_1'$} \put(6.6,2.2){$p_2$}
\put(9.7,2.2){$p_2'$} \put(6.3,3.4){$i$}          \put(10.15,3.7){$i'$}
\put(6.3,2.4){$j$}          \put(10.15,2.2){$j'$} \put(5,1){Fig. 1}
\end{picture}
\unitlength1cm
\centerline{\begin{picture}(9,3)
\put(1,2){\line(1,0){3}} \multiput(2,2)(1.5,0){2}{\vector(-1,0){.3}}
\put(1,.5){\line(1,0){3}} \multiput(1.7,.5)(1.5,0){2}{\vector(1,0){.3}}
\put(1,.7){$\r$} \put(4,.7){$\r'$}
\put(1,2.2){$\s$} \put(4,2.2){$\s'$}
\multiput(2.5,.65)(0,.3){5}{\line(0,-1){.2}}
\put(2.5,.5){\circle*{.1}} \put(2.5,2){\circle*{.1}}
\put(4.5,1.23){$-$}
\put(5,2){\line(1,0){3}} \multiput(6,2)(1.5,0){2}{\vector(-1,0){.3}}
\put(5,.5){\line(1,0){3}} \multiput(5.7,.5)(1.5,0){2}{\vector(1,0){.3}}
\put(4.8,.7){$\r$} \put(8.1,.7){$\r'$} \put(8.5,.7){;}
\put(4.8,2.2){$\s$} \put(8.1,2.2){$\s'$}
\multiput(6.5,.65)(0,.3){5}{\circle*{.15}}
\put(6.6,1.2){$K_{BR}$}
\put(6.5,.5){\circle*{.1}} \put(6.5,2){\circle*{.1}}
\put(4.2,0){2.a}
\end{picture}}


\input feynman

\newcommand{\ds}[4]{\global\seglength=1416 \global\gaplength=850
                    \drawline\scalar[#1\REG](#2,#3)[#4] }
\centerline{\begin{picture}(12000,8000)
\drawline\fermion[\E\REG](1000,7000)[10000]
\multiput(2200,7000)(7000,0){2}{\vector(-1,0){.3}}
\drawline\fermion[\E\REG](1000,1000)[10000]
\multiput(2700,1000)(7000,0){2}{\vector(1,0){.3}}
\drawline\gluon[\N\FLIPPEDCURLY](6000,1000)[7]
 \put(6000,7000){\circle*{250}}
\put(6000,1000){\circle*{250}}
\put(5000,-500){2.b}
\end{picture}
\begin{picture}(12000,8000)
\drawline\photon[\E\REG](3000,4000)[6]
\drawline\fermion[\NW\REG](\photonfrontx,\photonfronty)[2500]
\put(\photonfrontx,\photonfronty){\circle*{250}}
\put(\photonfrontx,\photonfronty){\vector(-1,1){1000}}
\drawline\fermion[\SW\REG](\photonfrontx,\photonfronty)[2500]
\put(\photonfrontx,\photonfronty){\vector(1,1){-800}}
\drawline\fermion[\NE\REG](\photonbackx,\photonbacky)[2500]
\put(\photonbackx,\photonbacky){\circle*{250}}
\put(\photonbackx,\photonbacky){\vector(-1,-1){-800}}
\drawline\fermion[\SE\REG](\photonbackx,\photonbacky)[2500]
\put(\photonbackx,\photonbacky){\vector(1,-1){1000}}
\put(5500,-500){2.c}
\end{picture}}
\centerline{Fig. 2}

\newpage

\begin{picture}(20000,18000)
\drawline\fermion[\E\REG](2000,15500)[16000]
\drawline\fermion[\E\REG](2000,2000)[16000]
\multiput(3500,15500)(12000,0){2}{\vector(-1,0){.3}}
\multiput(4000,2000)(12000,0){2}{\vector(1,0){.3}}
\ds{\N}{10000}{2000}{6}
\put(10000,15500){\circle*{250}}
\put(10000,2000){\circle*{250}}
\drawloop\gluon[\E 5](10000,11000)
\put(10000,11000){\circle*{250}}
\put(10000,\gluonbacky){\circle*{250}}
\put(9500,0){3.a}
\end{picture}
\begin{picture}(20000,18000)
\drawline\fermion[\E\REG](2000,15500)[16000]
\drawline\fermion[\E\REG](2000,2000)[16000]
\multiput(3500,15500)(12000,0){2}{\vector(-1,0){.3}}
\multiput(4000,2000)(12000,0){2}{\vector(1,0){.3}}
\ds{\S}{10000}{15500}{2}
\ds{\N}{10000}{2000}{2}
\put(10000,15500){\circle*{250}}
\put(10000,2000){\circle*{250}}
\drawloop\gluon[\E 5](10250,11000)
\drawloop\gluon[\W 5](9750,6125)
\put(9750,6125){\line(1,0){500}}
\put(9750,11000){\line(1,0){500}}
\put(10000,11000){\circle*{250}}
\put(10000,6125){\circle*{250}}
\put(9500,0){3.b}
\end{picture}

\begin{picture}(20000,18000)
\drawline\fermion[\E\REG](2000,15500)[16000]
\drawline\fermion[\E\REG](2000,2000)[16000]
\multiput(3500,15500)(12000,0){2}{\vector(-1,0){.3}}
\multiput(4000,2000)(12000,0){2}{\vector(1,0){.3}}
\ds{\S}{10000}{15500}{2}
\ds{\N}{10000}{2000}{2}
\put(10000,15500){\circle*{250}}
\put(10000,2000){\circle*{250}}
\put(10000,8750){\circle{6000}}
\put(8000,8850){\vector(0,1){.3}}
\put(12000,8650){\vector(0,-1){.3}}
\put(10000,10750){\circle*{250}}
\put(10000,6750){\circle*{250}}
\put(9500,0){3.c}
\end{picture}

\begin{picture}(20000,12000)
\drawline\fermion[\E\REG](2000,8000)[16000]
\drawline\fermion[\E\REG](2000,2000)[16000]
\multiput(3500,8000)(12000,0){2}{\vector(-1,0){.3}}
\multiput(4000,2000)(12000,0){2}{\vector(1,0){.3}}
\ds{\N}{10000}{2000}{3}
\put(10000,8000){\circle*{250}}
\put(10000,2000){\circle*{250}}
\drawloop\gluon[\N 5](7500,8000)
\put(7500,8000){\circle*{250}}
\put(\gluonbackx,\gluonbacky){\circle*{250}}
\put(19000,0){3.d}
\end{picture}
\begin{picture}(20000,12000)
\drawline\fermion[\E\REG](2000,8000)[16000]
\drawline\fermion[\E\REG](2000,2000)[16000]
\multiput(3500,8000)(12000,0){2}{\vector(-1,0){.3}}
\multiput(4000,2000)(12000,0){2}{\vector(1,0){.3}}
\ds{\N}{10000}{2000}{3}
\put(10000,8000){\circle*{250}}
\put(10000,2000){\circle*{250}}
\drawloop\gluon[\N 5](3000,8000)
\put(3000,8000){\circle*{250}}
\put(\gluonbackx,\gluonbacky){\circle*{250}}
\end{picture}

\begin{picture}(20000,18000)
\drawline\fermion[\E\REG](2000,2000)[16000]
 \multiput(4000,2000)(12000,0){2}{\vector(1,0){.3}}
\ds{\N}{10000}{2000}{3}
\put(\scalarbackx,\scalarbacky){\circle*{250}}
\drawline\gluon[\NE\REG](\scalarbackx,\scalarbacky)[6]
\drawline\scalar[\NW\REG](\scalarbackx,\scalarbacky)[4]
\drawline\fermion[\E\REG](2000,\gluonbacky)[16000]
\multiput(2500,\gluonbacky)(14500,0){2}{\vector(-1,0){.3}}
\put(\gluonbackx,\gluonbacky){\circle*{250}}
\put(3500,\gluonbacky){\circle*{250}}
\put(10000,2000){\circle*{250}}
\end{picture}
\begin{picture}(20000,18000)
\drawline\fermion[\E\REG](2000,2000)[16000]
\multiput(4000,2000)(12000,0){2}{\vector(1,0){.3}}
\ds{\N}{10000}{2000}{3}
\put(\scalarbackx,\scalarbacky){\circle*{250}}
\drawline\gluon[\NW\REG](\scalarbackx,\scalarbacky)[6]
\drawline\scalar[\NE\REG](\scalarbackx,\scalarbacky)[4]
\drawline\fermion[\E\REG](2000,\gluonbacky)[16000]
\multiput(2500,\gluonbacky)(14500,0){2}{\vector(-1,0){.3}}
\put(\gluonbackx,\gluonbacky){\circle*{250}}
\put(16500,\gluonbacky){\circle*{250}}
\put(10000,2000){\circle*{250}}
\put(-1500,0){3.e}
\end{picture}

 \begin{picture}(20000,18000)
\drawline\fermion[\E\REG](2000,2000)[16000]
 \multiput(4000,2000)(12000,0){2}{\vector(1,0){.3}}
\ds{\N}{10000}{2000}{3}
\put(\scalarbackx,\scalarbacky){\circle*{250}}
\drawline\gluon[\NW\REG](\scalarbackx,\scalarbacky)[6]
\put(\gluonbackx,\gluonbacky){\circle*{250}}
\drawline\gluon[\NE\REG](\scalarbackx,\scalarbacky)[6]
\drawline\fermion[\E\REG](2000,\gluonbacky)[16000]
\multiput(2500,\gluonbacky)(14500,0){2}{\vector(-1,0){.3}}
\put(\gluonbackx,\gluonbacky){\circle*{250}}
\put(10000,2000){\circle*{250}}
\put(9500,0){3.f}
\end{picture}
\begin{picture}(20000,18000)
\drawline\fermion[\E\REG](2000,2000)[16000]
\multiput(4000,2000)(12000,0){2}{\vector(1,0){.3}}
\drawline\photon[\NE\REG](6000,2000)[12]
\put(\photonbackx,\photonbacky){\circle*{250}}
\drawline\fermion[\E\REG](2000,\photonbacky)[16000]
\multiput(2500,\photonbacky)(14500,0){2}{\vector(-1,0){.3}}
\put(\photonbackx,2000){\circle*{250}}
\drawline\photon[\NW\REG](\photonbackx,2000)[12]
\put(\photonbackx,\photonbacky){\circle*{250}}
\put(6000,2000){\circle*{250}}
\put(9500,0){3.g}
\end{picture}

 \begin{picture}(20000,18000)
\drawline\fermion[\E\REG](2000,2000)[16000]
 \multiput(4000,2000)(12000,0){2}{\vector(1,0){.3}}
\ds{\N}{10000}{2000}{3}
\put(\scalarbackx,\scalarbacky){\circle*{250}}
\drawline\gluon[\NW\REG](\scalarbackx,\scalarbacky)[6]
\put(\gluonbackx,\gluonbacky){\circle*{250}}
\drawline\gluon[\NE\REG](\scalarbackx,\scalarbacky)[6]
\ds{\N}{\scalarbackx}{\scalarbacky}{3}
\drawline\fermion[\E\REG](2000,\gluonbacky)[16000]
\multiput(2500,\gluonbacky)(14500,0){2}{\vector(-1,0){.3}}
\put(\gluonbackx,\gluonbacky){\circle*{250}}
\put(10000,2000){\circle*{250}}
\put(10000,\fermionbacky){\circle*{250}}
\put(9500,0){3.h}
\end{picture}

\centerline{Fig. 3}

\newpage

\begin{picture}(20000,18000)
\drawline\fermion[\E\REG](2000,15500)[16000]
\drawline\fermion[\E\REG](2000,2000)[16000]
\multiput(3500,15500)(12000,0){2}{\vector(-1,0){.3}}
\multiput(4000,2000)(12000,0){2}{\vector(1,0){.3}}
\ds{\S}{10000}{15500}{6}
\put(10000,15500){\circle*{250}}
\put(10000,2000){\circle*{250}}
\drawloop\gluon[\E 5](10250,11000)
\drawloop\gluon[\W 5](9750,6125)
\put(9750,6125){\line(1,0){500}}
\put(9750,11000){\line(1,0){500}}
\put(10000,11000){\circle*{250}}
\put(10000,6125){\circle*{250}}
\put(9500,0){4.a}
\end{picture}
\begin{picture}(20000,18000)
\drawline\fermion[\E\REG](2000,15500)[16000]
\drawline\fermion[\E\REG](2000,2000)[16000]
\multiput(3500,15500)(12000,0){2}{\vector(-1,0){.3}}
\multiput(4000,2000)(12000,0){2}{\vector(1,0){.3}}
\ds{\S}{10000}{15500}{6}
\put(10000,15500){\circle*{250}}
\put(10000,2000){\circle*{250}}
\drawloop\gluon[\E 5](10000,8750)
\put(10000,\gluonbacky){\circle*{250}}
\drawloop\gluon[\W 5](10000,8750)
\put(10000,\gluonbacky){\circle*{250}}
\put(10000,8750){\circle*{250}}
\put(9500,0){4.b}
\end{picture}

\begin{picture}(20000,18000)
\drawline\fermion[\E\REG](2000,15500)[16000]
\drawline\fermion[\E\REG](2000,2000)[16000]
 \multiput(3500,15500)(12000,0){2}{\vector(-1,0){.3}}
\multiput(4000,2000)(12000,0){2}{\vector(1,0){.3}}
\ds{\S}{10000}{15500}{6}
\put(10000,15500){\circle*{250}}
\put(10000,2000){\circle*{250}}
\put(10000,9500){\line(1,0){250}}
\drawloop\gluon[\E 5](10250,9500)
\put(9750,\gluonbacky){\line(1,0){500}}
\put(10000,\gluonbacky){\circle*{250}}
\drawloop\gluon[\W 3](9750,\gluonbacky)
\drawline\gluon[\N \REG](\gluonbackx,\gluonbacky)[2]
\drawloop\gluon[\N 3](\gluonbackx,\gluonbacky)
\put(9750,\gluonbacky){\line(1,0){250}}
\put(10000,9500){\circle*{250}}
\put(10000,\gluonbacky){\circle*{250}}
\put(9500,0){4.c}
\end{picture}

\centerline{Fig. 4}


\newpage

\begin{picture}(20000,18000)
\drawline\fermion[\E\REG](2000,2000)[16000]
\multiput(3750,2000)(12750,0){2}{\vector(1,0){.3}}
\drawline\scalar[\N\REG](5000,2000)[6]
\drawline\fermion[\E\REG](2000,\scalarbacky)[16000]
\multiput(3750,\scalarbacky)(12750,0){2}{\vector(-1,0){.3}}
\put(\scalarbackx,\scalarbacky){\circle*{250}}
\put(5000,9000){\circle*{250}}
\drawline\gluon[\E\REG](5000,9000)[9]
\put(\gluonbackx,\gluonbacky){\circle*{250}}
\put(\gluonbackx,2000){\circle*{250}}
\drawline\scalar[\N\REG](\gluonbackx,2000)[6]
\put(\scalarbackx,\scalarbacky){\circle*{250}}
\put(5000,2000){\circle*{250}}
\put(9500,0){5.a}
\end{picture}
\begin{picture}(20000,18000)
\drawline\fermion[\E\REG](2000,2000)[16000]
\multiput(3750,2000)(12750,0){2}{\vector(1,0){.3}}
\drawline\gluon[\N\REG](5000,2000)[11]
\drawline\fermion[\E\REG](2000,\gluonbacky)[16000]
\multiput(3750,\gluonbacky)(12750,0){2}{\vector(-1,0){.3}}
\put(\gluonbackx,\gluonbacky){\circle*{250}}
\put(5000,8000){\circle*{250}}
\drawline\photon[\E\REG](5000,8000)[9]
\put(\photonbackx,\photonbacky){\circle*{250}}
\put(\photonbackx,2000){\circle*{250}}
\drawline\gluon[\N\REG](\photonbackx,2000)[11]
\put(\gluonbackx,\gluonbacky){\circle*{250}}
\put(5000,2000){\circle*{250}}
\put(9500,0){5.b}
\end{picture}

\begin{picture}(20000,18000)
\drawline\fermion[\E\REG](2000,2000)[16000]
\multiput(3750,2000)(12750,0){2}{\vector(1,0){.3}}
\drawline\photon[\N\REG](6000,2000)[6]
\newcounter{ax}
\newcounter{ay}
\setcounter{ax}{\photonbackx}
\setcounter{ay}{\photonbacky}
\drawline\photon[\E\REG](6000,\photonbacky)[8]
\newcounter{bx} \newcounter{by}
\setcounter{bx}{\photonbackx}
\setcounter{by}{\photonbacky}
\put(\photonfrontx,\photonfronty){\circle*{250}}
\drawline\photon[\N\REG](\arabic{bx},2000)[6]
\put(\photonbackx,\photonbacky){\circle*{250}}
\drawline\photon[\NW\REG](\arabic{bx},\arabic{by})[10]
\put(\photonbackx,\photonbacky){\circle*{250}}
\drawline\photon[\NE\REG](\arabic{ax},\arabic{ay})[10]
\drawline\fermion[\E\REG](2000,\photonbacky)[16000]
\multiput(3750,\photonbacky)(12750,0){2}{\vector(-1,0){.3}}
\put(\photonbackx,\photonbacky){\circle*{250}}
\put(\arabic{bx},2000){\circle*{250}}
\put(\photonbackx,\photonbacky){\circle*{250}}
\put(6000,2000){\circle*{250}}
\put(9500,0){5.c}
\end{picture}
\begin{picture}(20000,18000)
\drawline\fermion[\E\REG](2000,2000)[16000]
\multiput(3500,2000)(13500,0){2}{\vector(1,0){.3}}
\drawline\scalar[\NE\REG](4200,2000)[4]
\put(\scalarbackx,\scalarbacky){\circle*{250}}
\drawline\gluon[\NW\REG](\scalarbackx,\scalarbacky)[5]
\put(\gluonbackx,\gluonbacky){\circle*{250}}
\drawline\gluon[\NE\REG](\scalarbackx,\scalarbacky)[5]
\drawline\scalar[\SE\REG](\scalarbackx,\scalarbacky)[4]
\put(\scalarbackx,\scalarbacky){\circle*{250}}
\drawline\fermion[\E\REG](2000,\gluonbacky)[16000]
\multiput(3500,\gluonbacky)(13500,0){2}{\vector(-1,0){.3}}
\put(\gluonbackx,\gluonbacky){\circle*{250}}
\put(4200,2000){\circle*{250}}
\put(9500,0){5.d}
\end{picture}

\centerline{Fig. 5}

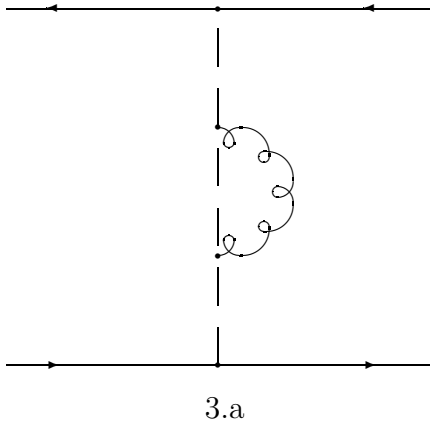
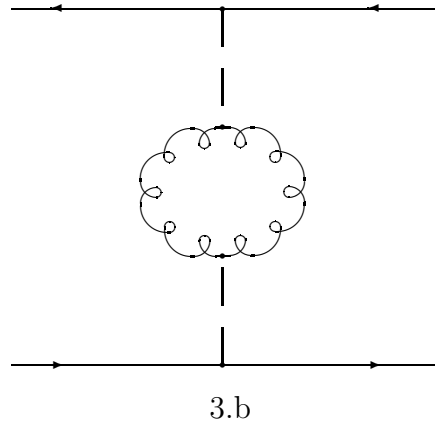
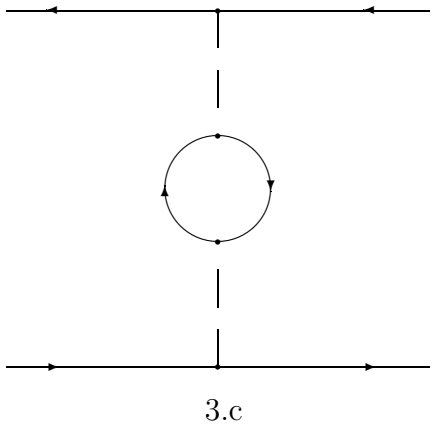
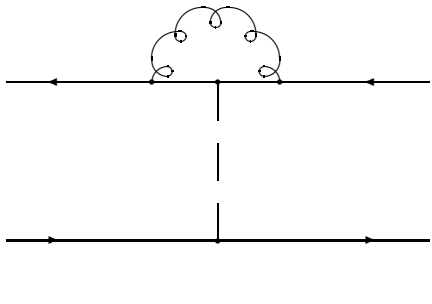
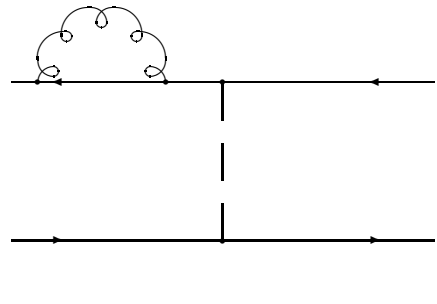
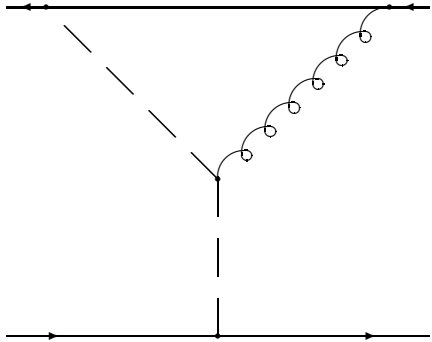
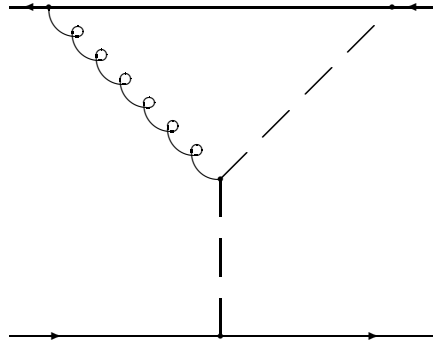
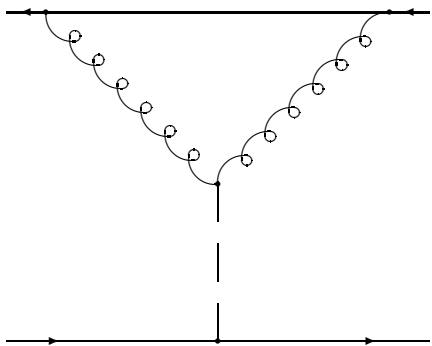
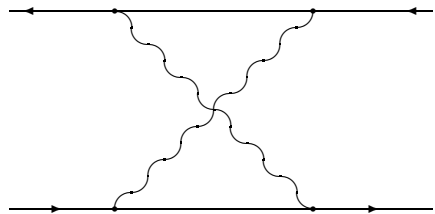
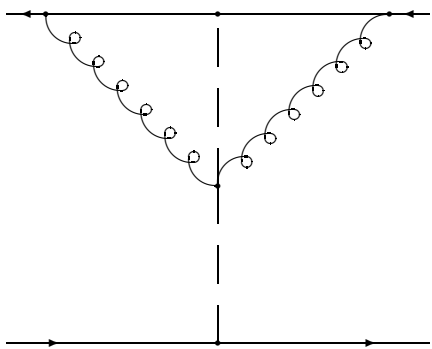
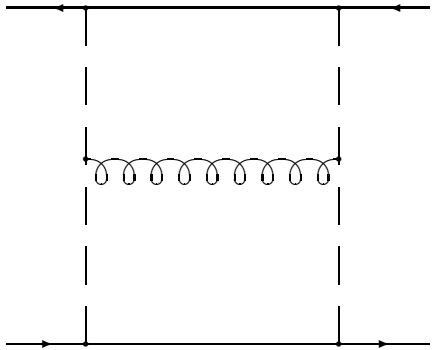
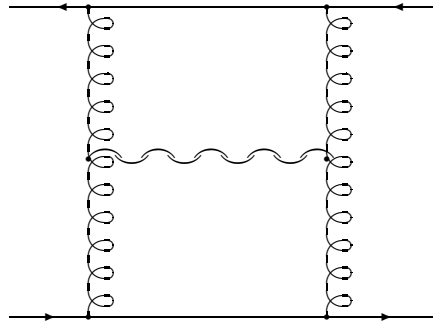
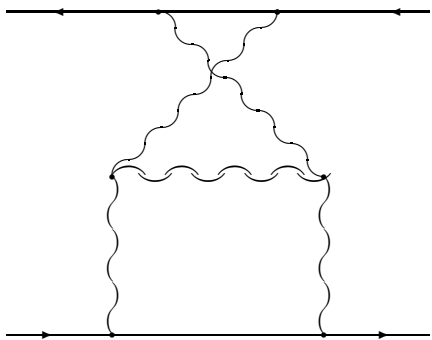
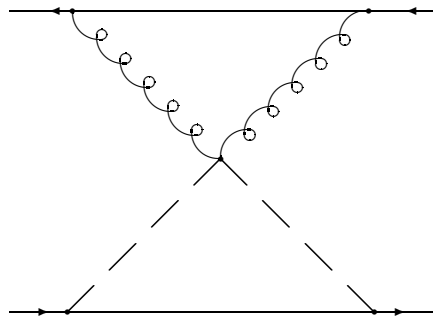
\end{document}